\documentclass[aps,twocolumn,amsmath,amssymb,floatfix,superscriptaddress,nofootinbib]{revtex4-1}

\usepackage{graphicx}
\usepackage{dcolumn}
\usepackage{bm}
\usepackage{latexsym}
\usepackage{color}
\usepackage{relsize}
\usepackage[colorlinks=true,linkcolor=blue,citecolor=blue]{hyperref}
\usepackage{wrapfig}

\newcommand{\lsc}{l_{\rm scatt}}

\begin{document}

\title{Physics of relativistic collisionless shocks:\\ III The
  suprathermal particles}

\author{Martin Lemoine}
\affiliation{Institut d'Astrophysique de Paris, CNRS -- Sorbonne Universit\'e, 98 bis boulevard Arago, F-75014 Paris}
\author{Guy Pelletier} 
\affiliation{UJF-Grenoble, CNRS-INSU, Institut de Plan\'etologie et d'Astrophysique de
Grenoble (IPAG), F-38041 Grenoble, France}
\author{Arno Vanthieghem}
\affiliation{Institut d'Astrophysique de Paris, CNRS -- Sorbonne Universit\'e, 98 bis boulevard Arago, F-75014 Paris}
\affiliation{Sorbonne Universit\'e, Institut Lagrange de Paris (ILP),
98 bis bvd Arago, F-75014 Paris, France}
\author{Laurent Gremillet}
\affiliation{CEA, DAM, DIF, F-91297 Arpajon, France}

\date{\today}

\begin{abstract}  
In this third paper of a series, we discuss the physics of the population of accelerated particles in the precursor of an unmagnetized, relativistic collisionless pair shock. In particular, we provide a theoretical estimate of their scattering length $\lsc(p)$ in the self-generated electromagnetic turbulence, as well as an estimate of their distribution function. We obtain $\lsc(p)\,\approx\,\gamma_{\rm p}\epsilon_B^{-1}(p/\gamma_\infty mc)^2c/\omega_{\rm p}$, with $p$ the particle momentum in the rest frame of the shock front, $\epsilon_B$ the strength parameter of the microturbulence, $\gamma_{\rm p}$ the Lorentz factor of the background plasma relative to the shock front and $\gamma_\infty$ its asymptotic value outside the precursor.  We compare this scattering length to large-scale PIC simulations and find good agreement for the various dependencies.
\end{abstract}

\maketitle

\section{Introduction}\label{sec:introd}
In recent decades, relativistic collisionless shock waves have emerged as outstanding phenomena of high energy astrophysics. As natural consequences of the powerful outflows associated with objects such as gamma-ray bursts, pulsar wind nebulae or active galactic nuclei, these shock waves seemingly dissipate with high efficiency the energy that drives them into powerlaws of relativistic charged particles, up to extreme energies, see {\it e.g.}~\cite{2012SSRv..173..309B,2015SSRv..191..519S,2017SSRv..207..319P} for reviews. By now, the basic mechanism of particle acceleration at shock waves is well understood in the test particle limit, in which one neglects the backreaction of these suprathermal particles on the shock and its environment, {\it e.g.}~\cite{1998PhRvL..80.3911B,2000ApJ...542..235K, 2001MNRAS.328..393A,2003ApJ...589L..73L,2005PhRvL..94k1102K, 2006ApJ...645L.129L,2006ApJ...650.1020N}. Much of the ongoing effort focuses on the nonlinear backreaction of accelerated particles on the magnetized turbulence, and how this turbulence controls in turn the acceleration and radiation processes. The acceleration of particles, the generation of turbulence and the structuring of the shock are indeed recognized as three inseparable aspects of collisionless shocks. 

For relativistic shocks, the level of magnetization of the unshocked medium is expected to be so low -- $\sigma\,\equiv\,(u_{\rm A}/c)^2\,\sim\,10^{-9}$ for the interstellar medium ($u_{\rm A}$ Alfv\'en four-velocity) -- that the magnetized turbulence is believed to be entirely generated by micro-instabilities acting in the shock precursor, which arise from the interpenetration of the beam of suprathermal particles and the unshocked (background) plasma~\cite{1999ApJ...526..697M,*1999ApJ...511..852G}. In other words, the suprathermal particles self-generate the magnetized turbulence in which they themselves scatter to be accelerated, and possibly also radiate at late times.

The present paper belongs to a series in which we explore the physics of unmagnetized relativistic pair shock fronts, using detailed analytical developments that we compare to dedicated particle-in-cell (PIC) simulations~\cite{L1}. In Paper~I~\cite{pap1}, we have argued that the microturbulence in the shock precursor is of a mostly magnetostatic nature in a frame noted $\mathcal{R}_{\rm w}$, which is found to move sub-relativistically with respect to the background plasma. In Paper~II~\cite{pap2}, we have developed a model of the heating and deceleration of the background plasma through its interaction with the microturbulence. Finally, Paper~IV~\cite{pap4} will address the evolution of the microturbulence in the shock precursor.

In this Paper~III, we focus on the physics of the suprathermal particles. In particular, we calculate their scattering length in the microturbulence and derive their distribution function in the precursor. As in the other papers of this series, we confront these theoretical predictions to dedicated PIC simulations. The scattering length $\lsc$ is a crucial quantity because it determines the time it takes to complete a Fermi cycle around the shock, hence the acceleration timescale, and hence the maximum energy of accelerated particles. Estimates of this acceleration timescale have been provided through test-particle simulations in a general model of the microturbulence~\cite{2013MNRAS.430.1280P} or through a direct measurement in long-timescale PIC simulations~\cite{2013ApJ...771...54S}, yet a clear analytical determination of $\lsc$ is still missing. This calculation is not straightforward, in particular because of nontrivial dependencies on the anisotropy of the turbulence spectrum~\cite{2007A&A...475...19A}. This length scale, defined in the shock rest frame, reveals a nontrivial scaling with $\gamma_\infty$, which plays an important phenomenological role as we argue in Sect.~\ref{sec:disc}. 

In order to model the backreaction of the suprathermal particles on the precursor, one needs to determine the evolution of their distribution function. In the following we carry out such a calculation in both the steady-state and time-dependent regimes, the latter being most relevant to PIC simulations.  This model also provides us a way to infer $l_{\rm scatt}$ from PIC simulations, which we compare with its theoretical value. Besides, those results help develop the scenario of heating and slowdown of the background plasma, presented in Paper II.
  
This paper is laid out as follows. Section~\ref{sec:setup} defines the main quantities used in this work; Sect.~\ref{sec:scatt} presents a theoretical calculation of the scattering length scale in a microturbulence in relativistic motion with respect to the shock front; Sect.~\ref{sec:suprat} discusses the properties of the suprathermal particle population, how it is distributed in the precursor, while clearly defining the stationary and time-dependent regimes; finally, Sect.~\ref{sec:disc} summarizes our results and draws some phenomenological consequences. We use Gaussian cgs units with $k_{\rm B}=c=1$.

\section{Physical and simulation setups}\label{sec:setup}
The general setup is as follows: we consider an unshocked background plasma  -- quantities indexed with subscript $_{\rm p}$ -- inflowing from $+\infty$ towards $-\infty$ along the $x-$axis (shock normal), in the (lab) frame in which the shock lies at rest at $x\,=\,0$, noted $\mathcal R_{\rm s}$. The Lorentz factor $\gamma_{\rm p}$ of the background plasma, before it enters the shock precursor, is written $\gamma_\infty$, its density $n_\infty$, etc. Quantities indexed with $_{\vert\rm w},\,_{\vert\rm p},\,_{\vert\rm d}$ are respectively defined in the turbulence frame $\mathcal R_{\rm w}$ (see~\cite{L1,pap1,pap2} and below), in the background plasma rest frame and in the downstream rest frame, which coincides with the reference frame of our PIC simulations. Quantities that are not indexed are by default considered in the shock frame, unless they are proper.

The shock precursor is defined as that region (of finite extent $\ell_{\rm prec}$) at positive values of $x$ in which the background plasma (indexed with $_{\rm p}$) interpenetrates a gas of suprathermal particles (indexed with $_{\rm b}$). The latter  correspond to particles that have been reflected on the shock, or accelerated through a Fermi-like process by multiple interactions around the shock front. As a result of the shock-crossing energy and momentum conserving equations in the fluid limit~\cite{1976PhFl...19.1130B}, the typical momentum of shock-heated particles is $\sim\,\gamma_\infty m$, and that of suprathermal particles is larger by a factor of the order of a few to ten. It is expected that these suprathermal particles exhibit a Maxwell-J\"uttner thermal distribution that turns over into a power-law of index $-s$ at large values of the momentum.

As discussed in various references, {\it e.g.}~\cite{1999ApJ...526..697M,*1999ApJ...511..852G,2004A&A...428..365W,2006ApJ...647.1250L, 2007A&A...475....1A,2007A&A...475...19A, 2010MNRAS.402..321L,2011MNRAS.417.1148L,2011ApJ...736..157R} for theoretical considerations, \cite{2007ApJ...668..974K,2008ApJ...673L..39S,2008ApJ...682L...5S, 2009ApJ...695L.189M,2009ApJ...693L.127K, 2009ApJ...698L..10N,2013ApJ...771...54S} for PIC simulations, the interpenetration of the beam of suprathermal particles and the background plasma gives rise to current filamentation instability (CFI), which generates an electromagnetic turbulence on plasma length scales $c/\omega_{\rm p}$~\cite{Weibel_1959,Davidson_1972,2004PhRvE..70d6401B, 2008PhRvL.100t5008B,2010PhPl...17l0501B,2010PhRvE..81c6402B}. In the unmagnetized limit, the dominant mode appears to be the transverse CFI, which forms current density filaments oriented along the shock normal, surrounded by (toroidal) magnetic fields and (radial) electric fields. In the precursor of a relativistic shock, the interaction between the suprathermal beam and the background plasma is strongly asymmetric. In the $\mathcal R_{\rm s}$ frame, the incoming background plasma is dense, (initially) cold and fast, moving with Lorentz factor $\gamma_{\rm p}$ and forming a tight beam in momentum space with opening angle $\lesssim\,1/\gamma_{\rm p}$, while the suprathermal particle population is tenuous, relativistically hot, slow and roughly isotropic. As discussed in previous papers of this series~\cite{L1,pap1,pap2}, there exists a particular frame $\mathcal R_{\rm w}$, in which the filamentation instability gives rise to essentially magnetostatic turbulence modes, which move at subrelativistic velocities relative to the background plasma. In this frame, the suprathermal particles are highly energetic, forming a beam with typical momentum $\gamma_\infty\gamma_{\rm p}$ and opening angle $\sim\,1/\gamma_{\rm p}$, while the background plasma is (at least initially) cold and nearly isotropic, drifting at subrelativistic velocity. Due to their relatively low momenta, the plasma particles are mainly trapped in the magnetic structures in $\mathcal{R}_{\rm w}$. By contrast, the suprathermal particles are scattered on scales much larger that the size of those structures, and so propagate in an essentially ballistic manner.

\begin{figure*}
\includegraphics[width=0.8\textwidth]{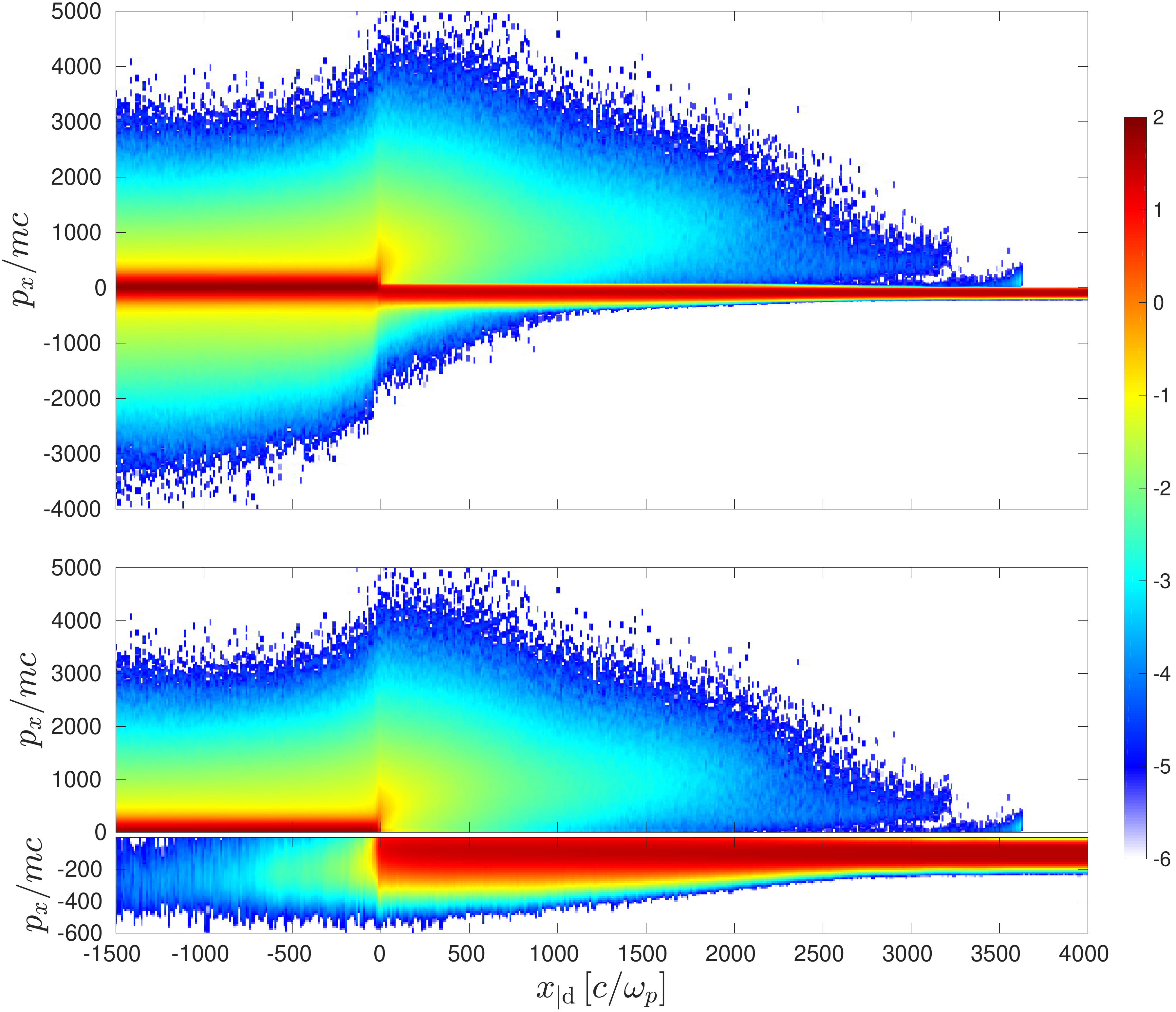}
 \caption{Density plot of the phase space $p_x/mc$ {\it vs} $x_{\vert\rm d}$ as measured in a PIC simulation with $\gamma_{\infty\vert\rm d}\,=\,100$. The color bar represents $\log({\rm d}N/{\rm d}x_{\vert\rm d}{\rm d}p_x)$ in arbitrary units, with $N$ the total number of particles. Top panel: total population of particles; middle panel: the beam suprathermal particle population, defined as those particles moving with $p_x>0$; bottom panel: the background plasma, {\it i.e.}, those particles moving with $p_x<0$ and having never experienced a turnaround in the turbulence. Note the different scales in each panel. 
  \label{fig:pspace} }
\end{figure*}

It is thus important to carefully distinguish these populations. In our numerical simulations, we define the background plasma particles as those particles that propagate with negative $x-$velocity continuously since their injection into the precursor, {\it i.e.}, their $x-$velocity has never changed sign. Oppositely, the suprathermal (or shock-reflected) particles are those moving with positive $x-$velocity, independently of the number of turn-arounds. Figure~\ref{fig:pspace} shows the phase space $p_x/m$ {\it vs} $x$ of the total and background plasma populations in one such simulation (with $\gamma_{\rm \infty \vert d} = 100$). 

Our 2D3V (2D in configuration space, 3D in momentum space) numerical simulations are performed using the finite-difference time-domain PIC code \textsc{calder}. They describe the self-consistent formation and evolution of an unmagnetized, relativistic collisionless shock wave in an electron-positron pair plasma. The shock is initialized by injecting the background plasma from the right-hand side and having it reflect specularly on a conducting wall, as in~\cite{2008ApJ...682L...5S}. The plasma is injected with proper temperature $T_\infty/m = 10^{-2}$  and Lorentz factor $\gamma_{\infty\vert\rm d}\,=\,10$ in the simulation frame, which coincides with the downstream rest frame. Hence the Lorentz factor of the background plasma relative to the shock front is $\gamma_\infty\,=\,17$. Both simulations initially employ 10 macro-particles per cell and per species, with a cell size $\Delta x \,=\,\Delta y\,=\, 0.1\,\omega_{\rm p}^{-1}$. The PIC code makes use of the Godfrey-Vay filtering algorithm~\cite{Godfrey_2014} and the Cole-Karkkainen finite difference field solver~\cite{Cole_1997a, Cole_2002, Karkkainen_2006} in order to mitigate the \v Cerenkov instability while preserving a large time step $\Delta t = 0.99 \Delta x $.

\begin{figure*}
\includegraphics[width=0.99\textwidth]{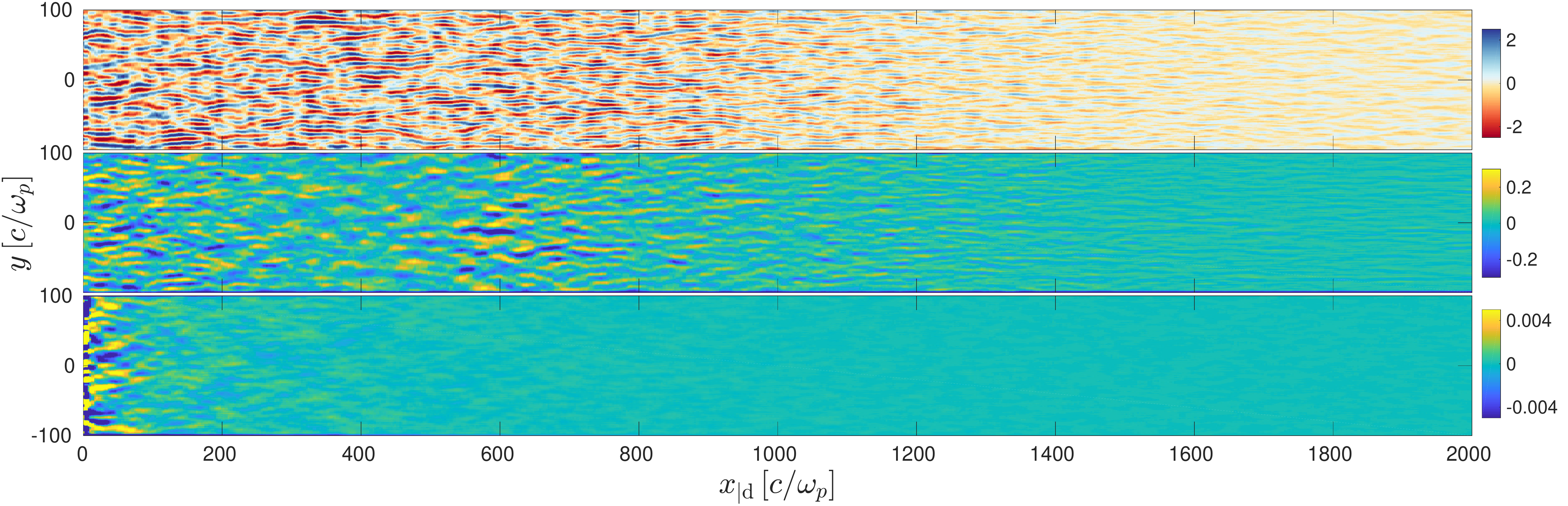}
 \caption{Top panel: density plot of $B_z$, in units of $\gamma_{\infty\vert\rm d}\epsilon_B^{1/2}$, from a PIC simulation with $\gamma_{\infty\vert\rm d}\,=\,100$, illustrating the filamentary structure of the turbulence over the precursor. Middle panel: density plot of $n_{\rm p}(x,y) - \left\langle n_{\rm p}\right\rangle_y$ (average taken over the transverse dimension), which reveals the filamentation pattern of the background plasma. Bottom panel: same for the suprathermal particles, on a scale enlarged by 50 to enhance their weak modulations.
  \label{fig:nofil} }
\end{figure*}

The asymmetry of the beam-plasma interaction in the precursor is clearly illustrated in Fig.~\ref{fig:nofil}, which displays the density variations of both populations as a result of the CFI in the $\gamma_{\rm \infty \vert d}\,=\,100$ case. The background plasma (middle) is seen to develop mildly nonlinear current filaments in response to the magnetic modulations (top), while the suprathermal particles (bottom) show very weak fluctuations only.

The present paper focuses on the characteristics of the suprathermal particle population. We first discuss the scattering length of these particles in the microturbulence -- thereby confirming, in particular, that it is much larger than the typical transverse size of a filament -- then consider their distribution function in the precursor. As shown in earlier papers of this series~\cite{L1,pap1,pap2}, a key quantity is the kinetic pressure $p_{\rm b}$, parametrized in terms of the incoming momentum flux through $\xi_{\rm b}\,\equiv\,p_{\rm b}/\left(\gamma_\infty^2 n_\infty m\right)$.

\section{Scattering length scale}\label{sec:scatt}

The physics of scattering of suprathermal particles in the downstream flow of an unmagnetized, relativistic collisionless shock is relatively easy to understand. As  PIC simulations have demonstrated~\cite{2008ApJ...682L...5S,2008ApJ...673L..39S, 2008ApJ...674..378C,2009ApJ...695L.189M, 2009ApJ...693L.127K,2011ApJ...726...75S,2013ApJ...771...54S},
the downstream turbulence is essentially magnetostatic in $\mathcal R_{\rm d}$ and distributed on short length scales $\lambda_{\delta B}\,\sim\,\mathcal O\left(10\,\omega_{\rm p}^{-1}\right)$. A particle thus suffers a typical deflection by an angle $\delta\theta\,\simeq\,\pm\lambda_{\delta B}/r_{\rm g}$ upon crossing a coherence length scale $\lambda_{\delta B}$, $r_{\rm g}$ denoting the gyration radius in the average magnetic field. There follows the pitch-angle scattering frequency $\nu_{\rm scatt}\,\simeq \,\delta \theta^2/\lambda_{\delta B}\,\simeq\,\lambda_{\delta B}/r_g^2$. Hence, the scattering length, {\it i.e.} the length beyond which the acquired deflection becomes of the order of unity, is $\lsc\,\simeq\,r_{\rm g}^2/\lambda_{\delta B}$. That $\lsc(p)\,\propto\,p^2$ is expected because the turbulence wavelength is much smaller than the gyroradius of accelerated particles, see~\cite{2009MNRAS.393..587P,2010MNRAS.402..321L,2011A&A...532A..68P,2013MNRAS.430.1280P}.

In the shock precursor, this scattering issue needs a careful analysis because: (i) the beam of suprathermal particles is strongly anisotropic in the $\mathcal R_{\rm w}$ frame in which the turbulence is of a mostly magnetic nature; (ii) this frame is in relativistic motion with respect to the $\mathcal R_{\rm s}$ shock rest frame; (iii) the turbulence itself is strongly anisotropic. 

The strong relativistic beaming of the suprathermal particles thus entails that their scattering frequency transforms nontrivially from $\mathcal{R}_{\rm w}$ to $\mathcal{R}_{\rm s}$, namely, $\nu_{\rm scatt } \simeq 4\gamma_{\rm w}^3 \nu_{\rm scatt \vert w}$ (see Sec.~\ref{sec:statdist}).  The anisotropy of the turbulence also plays an important role. For instance, describing the Weibel turbulence as a collection of infinitely long filaments oriented along the $x-$axis leads to the conservation of the conjugate canonical momentum associated to $\dot x$, thus precluding pitch-angle diffusion of the suprathermal particles~\cite{2007A&A...475...19A}. 

In the following, we first compute the correlation time of the random force experienced by a suprathermal particle, then its scattering length in the framework of an extended quasilinear theory, taking into account the anisotropy of the turbulence, its growth in the precursor and the relative motion between the $\mathcal R_{\rm w}$ frame and the shock front.

\subsection{The correlation time of the random force}\label{sec:tcorr}

From a statistical mechanics perspective, the correlation time of the random force that a particle suffers along its trajectory is a crucial quantity. It is defined as
\begin{equation}
t_{\rm corr}\,\equiv\,\frac{1}{\left\langle\bm{\delta}\mathbf{F}(t)\cdot\bm{\delta}\mathbf{F}(t)\right\rangle}\int_0^{+\infty}{\rm d}\tau\,\left\langle 
\bm{\delta}\mathbf{F}(t+\tau)\cdot \bm{\delta}\mathbf{F}(t)\right\rangle\,,
\label{eq:tc1}
\end{equation}
with $\bm{\delta}\mathbf{F}(t)\,=\,q\left(\bm{\delta}\mathbf{E}+\bm{\beta}\times\bm{\delta}\mathbf{B}\right)$ the random Lorentz force exerted at time $t$, at position $\bm{x}(t)$ of the particle. Formally, $t$ and $\tau$ in Eq.~(\ref{eq:tc1}) above represent the time in the lab frame along the particle trajectory; however, for suprathermal particles, it is a good approximation to use for $\bm{x}(t)$ a straight line trajectory, because the scattering length scale is expected to be much larger than the correlation length, as a result of the small-scale nature of the turbulence. 

In the $\mathcal R_{\rm s}$ frame, $\bm{\delta}\mathbf{E}\,=\,-\bm{\beta_{\rm w}}\times\bm{\delta}\mathbf{B}$, hence
\begin{equation}
\left\langle 
\bm{\delta}\mathbf{F}(t+\tau)\cdot \bm{\delta}\mathbf{F}(t)\right\rangle\,\propto\,\left\langle\bm{\delta}\mathbf{B}(t+\tau)\cdot\bm{\delta}\mathbf{B}(t)\right\rangle\,.
\label{eq:tc2a}
\end{equation}
The prefactor depends on quantities that vary slowly on $t_{\rm corr}$ timescales, which therefore drop out when taking the ratio in Eq.~(\ref{eq:tc1}). Using $\bm{\delta}\mathbf{B}\,=\,\gamma_{\rm w}\bm{\delta}\mathbf{B_{\vert\rm w}}$, with $\bm{\delta}\mathbf{B_{\vert\rm w}}$ defined in the $\mathcal R_{\rm w}$ frame, we decompose the latter in $\mathcal R_{\rm w}$ plane waves, to obtain
\begin{align}
t_{\rm corr}\,=\,&\frac{1}{\mathcal N}\int{\rm d}^3{k'}\nonumber\\\
&\quad\times\int_0^{+\infty}{\rm d}\tau\,e^{-2g\left[x(t+\tau)- x(t)\right]}\,e^{+i{k'}^\alpha\,{\Delta x_{\vert\rm w}}_\alpha}{{\mathcal S}_{\vert\rm w}}(\mathbf{k'})\,,
\label{eq:tc2}
\end{align}
with $\mathcal N\,\equiv\,\int {\rm d}\bm{k'}\,{{\mathcal S}_{\vert\rm w}}(\mathbf{k'})$, and $\Delta x_{\vert\rm w}^\alpha$ the displacement of the particle along its world line during the time interval $\Delta t$. We also defined the power spectrum through
\begin{equation}
\left\langle \bm{\delta}\mathbf{B_{\vert\rm w}}_{\mathbf{k'_1}}\cdot\bm{\delta}\mathbf{B_{\vert\rm w}}_{\mathbf{k'_2}}^\star\right\rangle\,=\,\delta\left(\mathbf{k'_1}-\mathbf{k'_2}\right)\,{{\mathcal S}_{\vert\rm w}}(\mathbf{k'_1})\,.
\label{eq:Sk}
\end{equation}
The turbulence is assumed stationary in $\mathcal{R}_{\rm s}$, and we have extracted from $\delta B$ its spatial profile, taken in the form $\propto \exp(-2gx)$. This choice means that the fluctuation spectrum $S_{\vert \rm w}$ is normalized to a value of $\langle B_{\rm w}^2 \rangle$ close to the shock front. Note that the spatial growth rate $g$, which is assumed independent of $\mathbf{k'}$, is expressed in $\mathcal{R}_{\rm s}$ while the plane-wave expansion is performed in $\mathcal{R}_{\rm w}$.

The above decomposition of the magnetic field into a slowly evolving envelope $\exp(-g\,x)$ times a plane wave decomposition should be understood as a simplified description of the turbulence in the precursor; it assumes, in particular, that the power spectrum is preserved throughout the precursor while the turbulence magnetic energy grows. Note also that the quantity ${k'}^\alpha\,{\Delta x_{\vert\rm w}}_\alpha$ is a Lorentz scalar, hence it can be expressed in any frame.

The quantity ${\Delta x_{\vert\rm w}}^\alpha$ represents the spacetime difference in $\mathcal R_{\rm w}$ between the trajectory at $t+\tau$ and that at $t$. If the particle is initially emitted with a pitch angle cosine $\mu$ along the shock normal, in the $\mathcal R_{\rm s}$ frame, then
\begin{eqnarray}
{\Delta x_{\vert\rm w}}^t&\,=\,&\gamma_{\rm w}(1-\mu\beta_{\rm w})\tau\,,\nonumber\\
{\Delta x_{\vert\rm w}}^x&\,=\,&\gamma_{\rm w}(\mu-\beta_{\rm w})\tau\,,\nonumber\\
{\Delta x_{\vert\rm w}}^\perp&\,=\,&(1-\mu^2)^{1/2}\tau\,.
\label{eq:tc3}
\end{eqnarray}
Note that, because it is expressed in $\mathcal R_{\rm s}$, $\mu$ is in principle determined at random between $0$ and $1$. In the following, we approximate $\beta_{\rm w}\,\rightarrow\,-1$ when possible. We also note that the combination $-\omega'{\Delta x_{\vert\rm w}}^t+k'_x{\Delta x_{\vert\rm w}}^x$ can be rewritten $\gamma_{\rm w}(1+\mu)\left(1-v_{\omega'}\right)\tau$, introducing the quantity $v_{\omega'}\,=\,\omega'/k'_x$; $\omega'$ is real by definition, since the growing part of the magnetic turbulence has been extracted previously. 

For the purely transverse mode of the CFI, $\omega'\,=\,0$. However, at finite $k_x'$ (yet $k_{x}'\,\ll\,k_\perp'$), the cold fluid dispersion relation of the CFI yields $\omega'\,=\,\beta_{\rm b\vert\rm w}k'_x\left[1+\mathcal O(\xi_{\rm b})\right]$~\cite{2013MNRAS.430.1280P}, with $\beta_{\rm b\vert\rm w}$ the velocity of the beam in the $\mathcal R_{\rm w}$ frame. Then $v_\omega'\,\approx\,1$, because $\beta_{\rm b\vert\rm w}\,\simeq\,1$ to an accuracy of order $1/\gamma_{\rm b\vert\rm w}^2\,\sim\,1/\gamma_\infty^4$. In a realistic shock precursor, where the filamentation instability reaches a nonlinear stage, and where oblique modes may contribute to shaping the turbulence, $\omega'$ may not obey the above relation, yet we retain the scaling $\omega'\,\propto\,k_x'$ and discuss the influence of $v_{\omega'}$ on the result.

We now define the resonance function
\begin{align}
\mathcal R_{\mathbf{k'}}&\,\equiv\,\int_0^{+\infty}{\rm d}\tau\,\, e^{-2g \mu \tau}\,e^{+i{k'}^\alpha{\Delta x_{\vert\rm w}}_\alpha}\nonumber\\
&\,=\, \begin{cases}\displaystyle{\pi\,\delta\left({k'}^\alpha{\Delta x_{\vert \rm w}}_\alpha/\tau\right)}  & \quad (g\,\rightarrow\,0) \\ 
& \\
\displaystyle{\frac{4 g \mu}{4 g^2\mu^2 + \left({k'}^\alpha{\Delta x_{\vert \rm w}}_\alpha/\tau\right)^2}}  & \quad (g\,\neq\,0)\end{cases}
\label{eq:tc4}
\end{align}

To keep the integrals analytically tractable, we approximate the spectrum with a constant patch in $\mathbf{k'}-$space, centered on $(0,\widehat{k'_\perp})$ with extension $(\pm\Delta k'_x,\pm\Delta k'_\perp)$, describing a mostly transverse instability as it should:
\begin{align}
{{\mathcal S}_{\vert\rm w}}(\mathbf{k'})\,\equiv\,&\frac{\langle{\delta B_{\vert\rm w}}^2\rangle}{4\,\Delta k'_\perp\,\Delta k'_x}\Theta\left[\left(k'_\perp - \widehat{k'_\perp}\right) - \Delta k'_\perp\right]\nonumber\\
&\quad\times
\Theta\left[\Delta k'_\perp - \left(k'_\perp - \widehat{k'_\perp}\right)  \right]\Theta\left[k'_x+\Delta k'_x\right]\nonumber\\
&\quad\times\Theta\left[\Delta k'_x - k'_x\right]\,.
\label{eq:tc5}
\end{align}

We note that $g\,=\,g_{\vert\rm w}/\gamma_{\rm w}$, since the growth time of the instability $g^{-1}$ in $\mathcal R_{\rm s}$ is $\gamma_{\rm w}$ times that experienced by the background plasma in $\mathcal R_{\rm w}$~\cite{pap4}. Furthermore, $g_{\vert\rm w}\,\ll\,\widehat{k'_\perp}$ for the filamentation instability, hence the real part $g\mu$ is a small quantity relative to ${k'}^\alpha{\Delta x_{\vert\rm w}}_\alpha/\tau$.

Consider first the limit $g\,\rightarrow\,0$. Then the integral in Eq.~(\ref{eq:tc2}) depends on the quantities $\Delta k'_x$, $\widehat{k'_\perp}$ and $\Delta k'_\perp$ characterizing the spectrum as well as on $(1-v_{\omega'})$. More specifically,
\begin{equation}
t_{\rm corr}\,=\,\begin{cases}
\displaystyle{0} & \displaystyle{{\rm if}\quad \Delta k'_x\,<\,\nu\,
\frac{\widehat{k'}_\perp-\Delta k'_\perp}{\gamma_{\rm w}(1-v_{\omega'})}}\\
 & \\
\displaystyle{\frac{\pi}{2}\frac{1}{\left(1-v_{\omega'}\right)}\frac{1}{\gamma_{\rm w}\Delta k'_x}} & \displaystyle{{\rm if}\quad \Delta k'_x\,>\,
\nu\,\frac{\widehat{k'}_\perp-\Delta k'_\perp}{\gamma_{\rm w}(1-v_{\omega'})}}\,,
\end{cases}
\label{eq:tc6}
\end{equation}
with $\nu=(1-\mu)^{1/2}/(1+\mu)^{1/2}\,\approx\,1$.
The importance of $v_{\omega'}$ is thus clear. If $v_{\omega'}$ is small compared to unity, then the condition specified on the rhs of the above equations amounts to whether $\Delta k_x'\,<\,\Delta k_\perp'/\gamma_{\rm w}$ or not, assuming $\widehat{k'_\perp}\,\sim\,\Delta k_\perp'$. In the limit $\gamma_{\rm w}\,\gg\,1$, the latter condition $\Delta k_x'\,>\,\Delta k_\perp'/\gamma_{\rm w}$ appears likely, hence one should expect $t_{\rm corr}\,\propto\,\left(\gamma_{\rm w}\Delta k'_x\right)^{-1}$. There follows the expected result that (in the limit of small-angle deflections) a nonvanishing  $\Delta k'_x$ is required to ensure a finite $t_{\rm corr}$, and therefore pitch-angle diffusion.

However, if $v_{\omega'}\,\simeq\,1-\epsilon$, as for the linear growth of the CFI, the former condition may hold. The spectrum is then such that it forbids the resonance of a particle with waves, preventing pitch-angle diffusion in this linearized limit. Obviously, any finite width to the dispersion relation, as characterized by the contribution in $g$ for instance, will lead to resonance broadening and permit pitch-angle diffusion. To first order in $g$, one obtains
\begin{equation}
t_{\rm corr}\,=\, 
\begin{cases}
\displaystyle{\alpha_1\,\frac{g_{\vert\rm w}}{\gamma_{\rm w}\Delta k'_\perp \widehat{k'_\perp}}} & 
\displaystyle{{\rm if}\quad \Delta k'_x\,<\,\nu\,
\frac{\widehat{k'}_\perp-\Delta k'_\perp}{\gamma_{\rm w}(1-v_{\omega'})}}\\
 & \\
\displaystyle{\alpha_2\frac{1}{\left(1-v_{\omega'}\right)}\frac{1}{\gamma_{\rm w}\Delta k'_x}} & \displaystyle{{\rm if}\quad \Delta k'_x\,>\,
\nu\,\frac{\widehat{k'}_\perp-\Delta k'_\perp}{\gamma_{\rm w}(1-v_{\omega'})}}
\end{cases}
\label{eq:tc6b}
\end{equation}
with $\alpha_1\,=\,\mu\ln\left[(\widehat{k'_\perp}+\Delta k'_\perp)/(\widehat{k'_\perp}-\Delta k'_\perp)\right]/(1-\mu^2)\,\approx\,1$ and $\alpha_2\,=\, \pi \left[1+\mathcal O\left(g'/\widehat{k'_\perp}\,\Delta k'_x/\Delta k'_\perp \right)\right]/2\,\approx\,1$.
We are particularly interested in the dependence of $t_{\rm corr}$ on $\gamma_{\rm w}$ and, interestingly, both limits leads to $t_{\rm corr}\,\propto\,(\gamma_{\rm w}\widehat{k'_\perp})^{-1}$. The prefactor can be smaller or larger than unity, depending on which limit applies. Assuming for instance $\Delta k'_\perp\,\approx\,\widehat{k'_\perp}$, the first limit implies $\gamma_{\rm w}\widehat{k'_\perp} t_{\rm corr}\,\sim\,\mathcal O\left(g_{\vert\rm w}/  \widehat{k'_\perp}\right)$, which is typically an order of magnitude smaller than unity in the linear growth phase of the CFI in the precursor of a relativistic shock. The second limit yields $\gamma_{\rm w}\widehat{k'_\perp} t_{\rm corr}\,\sim\, \mathcal O\left(\widehat{k'_\perp}/\Delta k'_x\right)$, which is typically expected to be somewhat larger than unity. We will compare this prediction to measurements made in PIC simulations in the following.

Once the correlation time of the random force is known, one can estimate the scattering length (in $\mathcal R_{\rm s}$) by noting that, over $t_{\rm corr}$, the particle suffers a deflection of the order of $\pm ct_{\rm corr}/r_{\rm g}$, so that $\lsc\,\approx\,r_{\rm g}^2/(c t_{\rm corr})$. 

\subsection{Quasilinear estimate of $\lsc$}
We now carry out a quasilinear calculation of the pitch angle diffusion coefficient of suprathermal particles (with $\gamma_{\vert\rm p} \gg 1$) in the shock rest frame. In order to keep track of conserved quantities in the possible limit of time-independent or $x-$independent turbulence (describing infinitely long filaments), we rely on a Hamiltonian formalism for the equations of motion. We first note that, for suprathermal particles, the canonical conjugate momentum $\pi^\alpha\,\equiv\,p^\alpha\, +\, q\, \delta A^\alpha$ coincides with the momentum $p^\alpha$ to a small error of order $\lambda/r_{\rm g}\,\ll\,1$, because the four-vector potential $\vert \bm{\delta}\mathbf{A}\vert\,\sim\,\lambda \delta B$ in order of magnitude. We thus use the approximation
\begin{equation}
  \mu\,\simeq\,\frac{\pi^x}{\pi^t}
  \label{eq:muapp}
\end{equation}
to describe the evolution of the pitch angle cosine $\mu\,\equiv\,p^x/p$ in the shock frame, as
\begin{equation}
  \frac{{\rm d}\mu}{{\rm
      d}s}\,\simeq\,\frac{1}{{\pi^{t}}^2}\left(\pi^t\frac{{\rm
      d}\pi^x}{{\rm d}s}-\pi^x\frac{{\rm
      d}\pi^t}{{\rm d}s}\right)\,,
\label{eq:muapp2}
 \end{equation}
with the Hamilton equation 
\begin{equation}
  \frac{{\rm d}\pi^\alpha}{{\rm d}s}\,=\,q\left(\pi_\beta - q \,\delta A_\beta\right)\frac{\partial A^\beta}{\partial x_\alpha}\,.
\label{eq:muapp3}
\end{equation}
In the above two equations, the conservation of $\mu$ along the particle trajectory indexed by the affine coordinate $s$ is manifest if the turbulence is both time and $x-$independent. Under standard assumptions, if neither of these conditions holds, the pitch angle may start to diffuse; then, the error associated to Eq.~(\ref{eq:muapp}) is bound to decrease in time, so that our approximation will become more and more accurate. In effect, this error is bounded by the range of variation of the four-vector potential: $\Delta \mu\,\sim\,\vert\Delta \delta A\vert/\pi^t$; it thus remains fixed in time while the r.m.s. of the pitch angle cosine distribution increases through diffusion.

To simplify the notations, all unprimed variables are understood to be defined in $\mathcal R_{\rm s}$ in this section, while primed variables are defined in $\mathcal R_{\rm w}$. Substituting Eq.~(\ref{eq:muapp3}) in (\ref{eq:muapp2}), and using (\ref{eq:muapp}) gives
\begin{align}
  \Delta\mu(t)\,\simeq\,&q\int_{0}^{t}\frac{{\rm d}\tau}{p}\,
  \frac{1}{\pi^{t}}\left(\partial_x+\mu\partial_t\right)\delta A'^\alpha p'_\alpha\,.
\end{align}
Here, an extra factor of $1/p$ has appeared because ${\rm d}s\,\equiv\,{\rm d}\tau/p$, with ${\rm d}\tau$ a time interval defined in $\mathcal R_{\rm s}$; furthermore, ${\delta A'}^\alpha$, ${p'}_\alpha$ are now primed variables. The partial derivatives are more conveniently expressed in terms of primed partial derivatives, and ${A'}^\alpha$ can be decomposed in plane waves with polarization four-vectors ${{e'}_{\mathbf{k'}}}^\alpha$:
\begin{equation}
  {\delta A'}^\alpha\,=\,e^{-g\,x}\int\frac{{\rm
      d}^3k'}{(2\pi)^3}\,{{e'}_{\mathbf{k'}}}^\alpha\,e^{i{k'}_\mu\,
    {x'}^\mu}\,,
\end{equation}
where, as in Sec.~\ref{sec:tcorr}, the spatial $x-$profile of ${\delta A'}^\alpha$ has been extracted from the plane wave decomposition. Let us stress again that $g$ represents the growth length scale in the shock rest frame, and that it does not depend on $\mathbf{k'}$.

Consequently,
\begin{align}
  \Delta\mu(t)\,\simeq\,&\frac{e}{p^2}\int_{0}^{t}{\rm d}\tau\frac{{\rm d}^3k'}{(2\pi)^3}\,
  \gamma_{\rm w}\left[i(1-\beta_{\rm w}\mu)k_x' - i(\mu-\beta_{\rm w})\omega'\right]
\nonumber\\
&\quad\quad\times\,{p'}_\alpha\, {{e'}_{\mathbf{k'}}}^\alpha\,e^{-g\,x}\,
  e^{ik'_\mu\, x'^\mu} \,, 
\end{align}
where it is understood that $\omega'$ is real, since the growing part has been extracted previously.

The statistical properties of the microturbulence in the $\mathcal R_{\rm w}$ 
can be approximated through the correlators
\begin{equation}
  \left\langle {{e'}_{\mathbf{k'_1}}}^\alpha\,
  {{{e'}_{\mathbf{k'_2}}}^\beta}^\star\right\rangle\,=\,(2\pi)^3\,
  \delta\left(\mathbf{k_1'}-\mathbf{k_2'}\right){\mathcal S_{\mathbf{k'_1}}}^{\alpha\beta}\,.
\end{equation}
In particular, for an anisotropic axisymmetric configuration with
$k'_\perp\,>\,k'_x$, we can set
\begin{equation}
  {\mathcal S_{\mathbf{k'}}}^{\alpha\beta}\,=\,{\mathcal S_\parallel} \delta^\alpha_{x'}\delta^\beta_{x'} + {\mathcal S_\perp}\left(\delta^\alpha_{y'}\delta^\beta_{y'} + \delta^\alpha_{z'}\delta^\beta_{z'} \right)\,.
\end{equation}
The case of $S_\parallel (\mathbf{k'}) > S_\perp(\mathbf{k'})$ corresponds to filaments elongated along $x$, as considered here. The power spectrum $S_\parallel$ of $\mathbf{\delta A'}$ is related to the power spectrum $S_{\vert \rm w})$ of $\mathbf{\delta B'}$, defined in Eq. (\ref{eq:Sk}), through $\mathcal S_{\vert\rm w}(\mathbf{k'})\,=\, k_\perp'^2 \mathcal S_\parallel$, since
\begin{align}
  \langle \delta B'^2\rangle&\,=\,\int\frac{{\rm d}^3k'_1}{(2\pi)^3}\frac{{\rm
      d}^3k'_2}{(2\pi)^3} \, \left\langle
  \left(\mathbf{k_1'}\times\mathbf{e'_{k_1'}}\right)\cdot
  \left(\mathbf{k_2'}\times\mathbf
  {{e'_{\mathbf{k_2'}}}^\star}\right)\right\rangle\nonumber\\
&\,\simeq\,\int\frac{{\rm
      d}^2k_\perp'{\rm d}k_x'}{(2\pi)^3}\,k_\perp'^2\,\mathcal S_\parallel\,.
\end{align}

We now approximate the trajectory as rectilinear, as in Eq.~(\ref{eq:tc3}) and evaluate the evolution of pitch angle cosine over a time interval $\Delta t$ assumed much larger than the coherence time of the electromagnetic force $t_{\rm corr}$:
\begin{eqnarray}
  \left\langle\Delta\mu^2\right\rangle&\,\simeq\,&
  \Delta t \frac{e^2 \,p'^2_x}{p^4}\gamma_{\rm w}^2
  \int\frac{{\rm d}^3k'}{(2\pi)^3}\nonumber\\
&&\quad\quad\times\left[(1-\beta_{\rm w}\mu)k_x' -
    (\mu-\beta_{\rm w})\omega'\right]^2\,\mathcal R_{\mathbf{k'}}\,
 \mathcal S_\parallel\,,\nonumber\\
&&
\end{eqnarray}
where the response function $\mathcal R_{\mathbf{k'}}$ defined in Eq.~(\ref{eq:tc4}) appears through the time integration. We have approximated ${p'}_\alpha\, {p'}_\beta\langle
{{e'}_{\mathbf{k_1'}}}^\alpha\,{{{e'}_{\mathbf{k_2'}}}^\beta}^\star\rangle\,\simeq\,p_x'^2
\mathcal S_\parallel\,(2\pi)^3\,\delta\left(\mathbf{k_1'}-\mathbf{k_2'}\right)$ because $\mathcal S_\parallel\,>\,S_\perp$ and $p_x'/p_\perp'\,\sim\,\mathcal O(\gamma_{\rm w})$.

Everywhere in the integral, we can take the simplifying limit $\beta_{\rm w}\,\simeq\,-1$. We also note that $\langle \delta B'^2\rangle\,=\,\langle\delta B^2\rangle/\gamma_{\rm w}^2$, $\langle\delta B^2\rangle$ denoting the rms measured in the shock frame $\mathcal R_{\rm s}$. To compute the above integral, we use the same power spectrum as in Sec.~\ref{sec:tcorr} and we pay attention to the lowest order term in $g'/\widehat{k'_\perp}$. From the definition $\nu_{\rm scatt}\,=\,\left\langle\Delta \mu^2\right\rangle/2\Delta t$, we eventually obtain
\begin{equation}
\nu_{\rm scatt}(p)\,=\,\frac{e^2\langle\delta B^2\rangle}{p^2}\,
\begin{cases}
\displaystyle{\alpha_3\,\frac{g'}{\gamma_{\rm w}\Delta k'_\perp \widehat{k'_\perp}}} & 
\left[\Delta k'_x\,\lesssim\,\frac{\widehat{k'_\perp}}{\gamma_{\rm w}(1-v_{\omega'})}\right]\\
 & \\
\displaystyle{\alpha_4\,\frac{1}{\gamma_{\rm w}(1-v_{\omega'})\Delta k'_x}} & 
\left[\Delta k'_x\,\gtrsim\,\frac{\widehat{k'_\perp}}{\gamma_{\rm w}(1-v_{\omega'})}\right]
\end{cases}
\label{eq:tc6c}
\end{equation}
with $\alpha_3\,=\,\mu(1+\mu)^2\,\ln\left[(\widehat{k'_\perp}+\Delta k'_\perp)/(\widehat{k'_\perp}-\Delta k'_\perp)\right] \,\approx\,1$ and $\alpha_4\,=\, \pi(1-\mu)(1+\mu)^2/2\,\approx\,1$.
In both limits, one can verify that, to within a factor of the order of unity, $\nu_{\rm scatt}\,\simeq\,t_{\rm corr}/r_{\rm g}^2$, with $r_{\rm g}\,=\,p/(e\langle\delta B^2\rangle^{1/2})$, as expected.

To encompass both limits, we write:
\begin{equation}
\nu_{\rm scatt}\,\simeq\, \frac{e^2 \langle\delta B^2\rangle}{p^2}\frac{1}{\gamma_{\rm w}\,\overline{k'}}\,,
\label{eq:nus1}
\end{equation}
where $\overline{k'}$ is a wavenumber of approximate value $\widehat{k'}_\perp^2/g'$ if the $g'\rightarrow0$ resonance with waves cannot be satisfied (corresponding to the first limit) or $\Delta k'_x$ in the opposite case. Up to a numerical prefactor, which can be as large as an order of magnitude or so, we will assume in the following that $\overline{k'}\,\sim\,\omega_{\rm p}$, so that the above leads to our estimate for the scattering length $\lsc\,=\,\nu_{\rm scatt}^{-1}$:
\begin{equation}
 \lsc(p)\,\sim\,\gamma_{\rm p}\,\epsilon_B^{-1}\,
  \left(\frac{p}{\gamma_\infty m}\right)^2\,\omega_{\rm p}^{-1}\,,
\label{eq:lscfin}
\end{equation}
with $\epsilon_B \,=\,\langle\delta B^2\rangle/\left(4\pi\gamma_\infty^2 n m\right)$. Here, we have used the background plasma Lorentz factor $\gamma_{\rm p}$ as a proxy for $\gamma_{\rm w}$, which represents a good approximation, see Paper I~\cite{pap1}.

The above estimate of the scattering length can be understood in a simpler way if one omits the anisotropy of the turbulence. Consider a magnetostatic turbulence with typical wavenumber $k'$ (in $\mathcal R_{\rm w}$). For beam particles of Lorentz factor $\gamma'$ in $\mathcal R_{\rm w}$, we have $k' r'_{\rm g}\,\simeq\, (k'/\omega_{\rm p})\epsilon_B^{-1/2}\gamma_\infty \gamma'/\gamma_{\rm w}\,\gg\,1$. Therefore the particles suffer small-angle interactions each time they cross a coherence length $\sim k'^{-1}$ of the microturbulence; the correlation time of the force thus reads $\sim k'^{-1}$, and the angular scattering frequency $\nu_{\rm s}'\,\sim\,k'/(k'r'_{\rm g})^2\,\sim\,
\omega_{\rm p}(k'/\omega_{\rm p})^{-1}\epsilon_B\gamma'^{-2}\gamma_\infty^2/\gamma_{\rm w}^2$. This scattering frequency is defined in $\mathcal R_{\rm w}$, and to convert it to $\mathcal R_{\rm s}$, one needs to multiply it by $\gamma_{\rm w}^3$ -- see Eq.~(\ref{eq:nunup}) further below -- while expressing $\gamma'\,\simeq\,\gamma_{\rm w}\gamma$. This gives $\nu_{\rm scatt}\,\sim\,\omega_{\rm p}(k'/\omega_{\rm p})^{-1}\epsilon_B\gamma_{\rm w}^{-1}(\gamma_\infty m/p)^2$, as obtained above. The origin of the $\gamma_{\rm w}$ factor in the scattering length thus results from the motion of $\mathcal R_{\rm w}$  relative to $\mathcal R_{\rm s}$, not from the anisotropy. The latter rather introduces the various possible values of $\overline{k'}$.

\section{Distribution in the precursor}\label{sec:suprat}
In astrophysical sources, it is generally expected that, on the time scale that observations probe, the shock wave and the acceleration process have reached a steady state. This means, in particular, that leptons and hadrons have been accelerated up to their respective maximal energies, which are determined through the competition between the energy loss time scale or the age of the source and the characteristic acceleration time scale. Most PIC simulations, however, do not reach a steady state, because particle acceleration, when effective, appears as an unbounded process on the simulation time scale. To make a proper comparison between theoretical predictions and PIC simulations, or to extrapolate the results of these simulations to astrophysical objects, it thus becomes important to distinguish which regime, steady state or not, applies.

To see this, consider the transport of an accelerated particle of Lorentz factor $\gamma\,>\,\gamma_{\infty}$ in the precursor, as viewed in the shock rest frame. The penetration length scale of this particle is controled by its scattering length scale $\lsc(\gamma)$, so that, on general grounds, one expects the distribution function of those particles ${\rm d}N/{\rm d}\gamma{\rm d}x$ to fall off exponentially on length scales $x\,\gtrsim\,\lsc(\gamma)$. Hence, at a distance $x$, one typically finds particles with a Lorentz factor $\gamma$ such that $\lsc(\gamma)\,\sim\,x$. Since $\lsc(\gamma)$ is a growing function of $\gamma$, the larger its $\gamma$, the further away from the shock front the particle can propagate. Our particle of Lorentz factor $\gamma$ moves in a ballistic manner on timescales $t\,\ll\,\lsc$, but it starts to diffuse in the turbulence on timescales $t\,\gg\,\lsc$. Once the PIC simulation has reached a duration $t_{\rm max\vert d}$ such that it exceeds the acceleration time scale to produce particles of Lorentz factor $\gamma$, and such that the precursor extends well beyond the maximal distance where such particles can be found, one does not expect further evolution of the distribution function for those particles beyond time $t_{\rm max\vert d}$, hence the PIC simulation has reached a steady state for that Lorentz factor.

As the precursor extends at velocity $\simeq\,c$, and because the acceleration timescale is of the order of $\lsc$ in the shock frame, those two conditions amount to the same: $t_{\rm max\vert d}\,\gg\,\lsc(\gamma)$. Then, the region close to the shock front with $x\,<\,\lsc(\gamma)$, populated by particles of Lorentz factor $\lesssim\,\gamma$ can be considered in steady state, while particles of larger Lorentz factor $\gamma$ or in regions further away from the shock front evolve in a time-dependent manner as the simulation runs. In the following, we thus distinguish between these two limits.

\subsection{Stationary state}\label{sec:statdist}
To derive the distribution function of accelerated particles in the steady state, one can use the kinetic equation given in Paper~II~\cite{pap2}, which describes the evolution of the distribution function in a mixed coordinate system, with space variables given in the shock rest frame, and momentum variables expressed in the Weibel frame, in which the turbulence is mostly magnetostatic. We simplify further this equation by neglecting the inertial force terms, corresponding to the limit in which the scattering length of the accelerated particles is much larger than the transition scale of the shock, over which significant slowdown of the Weibel frame occurs. This simplification allows us to express analytically the approximate suprathermal particle distribution function, but it prevents a detailed comparison with PIC simulations in regions in which the background plasma effectively slows down. Unfortunately, a closed-form solution which takes this deceleration into account does not seem at hand. Note that the deceleration enters both through the inertial correction, through the presence of $\gamma_{\rm w}$ in the equation, as well as through the dependence of the scattering frequency over $\gamma_{\rm w}$. In a 3D momentum space, the equation takes the form
\begin{equation}
  \gamma_{\rm w}\left(\beta_{\rm w}+\mu_{\vert\rm w}\right)\frac{\partial}{\partial x} f_{\rm b}\,=
  \,\frac{1}{2}\frac{\partial}{\partial {\mu_{\vert\rm w}}}\left[\left(1-\mu_{\vert\rm w}^2\right)\nu_{\rm scatt \vert\rm w}\frac{\partial}{\partial {\mu_{\vert\rm w}}}f_{\rm b}\right]\,.
  \label{eq:fb0}
\end{equation}
This equation has been previously solved in full generality in Ref.~\cite{2000ApJ...542..235K} through an expansion in eigenfunctions, in the limit $\gamma_{\rm w}\,\rightarrow\,\gamma_\infty$ and $\nu_{\rm scatt \vert\rm w}$ independent of $p$. Here we obtain an approximate solution including the dependence of $\nu_{\rm scatt \vert\rm w}$ on $p$.

This equation assumes $p_{\vert\rm w}^t\,\simeq\,p_{\vert\rm w}$, and $\mu_{\vert\rm w}=p_{\vert\rm w}^x/p_{\vert\rm w}$ represents the cosine of the angle of the particle momentum with the shock normal. Finally, the right-hand side models the stochastic force felt by the particle in the magnetostatic turbulence, as characterized by a scattering frequency $\nu_{\rm scatt\vert  w}$ (``Weibel frame''). To simplify it further, we neglect the dependence of $\nu_{\rm scatt\vert  w}$ on $\mu_{\vert\rm w}$ and use the small-angle approximation $\mu_{\vert\rm w}\,\simeq\,1-\theta_{\vert\rm w}^2/2$.  The latter is a good approximation, because $\vert1-\mu_{\vert\rm w}\vert\,\lesssim\,1/\gamma_{\rm w}\,\ll\,1$ for ultrarelativistic particles traveling upstream. The above equation can be then rewritten as
\begin{align}
 & \left(1-\gamma_{\rm w}^2\theta_{\vert\rm w}^2\right)\frac{\partial}{\partial {\nu_{\rm scatt\vert  w}x}}f_{\rm b}\,=\,\nonumber\\
&\quad\quad \gamma_{\rm w}^3\frac{1}{\gamma_{\rm w}\theta_{\vert\rm w}}\frac{\partial}{\partial \left({\gamma_{\rm w}\theta_{\vert\rm w}}\right)}\left[\gamma_{\rm w} \theta_{\vert\rm w}\frac{\partial}{\partial \left({\gamma_{\rm w}\theta_{\vert\rm w}}\right)}f_{\rm b}\right]\,.
  \label{eq:fb1}
\end{align}
Assuming separation of variables,
\begin{equation}
  f_{\rm b}\,\simeq\,\exp\left[-k \nu_{\rm scatt\vert  w}(p_{\vert\rm w}) x\right]h_k\left(\gamma_{\rm w}\theta_{\vert\rm w}\right)\,,
  \label{eq:fb2}
\end{equation}
we obtain the simple solution, with $k\,=\,4\gamma_{\rm w}^3$,
\begin{equation}
  f_{\rm b}(x,p_{\vert\rm w},\mu_{\vert\rm w})\,\simeq\,C(p_{\vert\rm w})\, e^{-4\gamma_{\rm w}^3\nu_{\rm scatt\vert  w}(p_{\vert\rm w}) x -\gamma_{\rm w}^2\theta_{\vert\rm w}^2}\,,
  \label{eq:fb3}
\end{equation}
where $C(p_{\vert\rm w})$ is a function of $p_{\vert\rm w}$. This function can be determined by matching, on the shock front, the full solution in the upstream half-plane with that in the downstream half-plane, for a proper choise of boundary conditions, see {\it e.g.}~\cite{2000ApJ...542..235K}. This procedure then determines  $C(p_{\vert\rm w})$ as a power-law of index $s+2$.
The relationship between $\nu_{\rm scatt\vert  w}(p_{\vert\rm w})$ and the scattering length $\lsc(p)$, which we have defined earlier in the shock frame, is not trivial because of the relativistic motion of the microturbulence frame $\mathcal R_{\rm w}$ with respect to the shock front. The scattering length $l_{\rm scatt}$ is indeed defined as the length scale over which suprathermal particles are deflected by an angle of order unity in the $\mathcal R_{\rm s}$ frame, in a time interval $\Delta t\,=\,\lsc$; however, in the $\mathcal R_{\rm w}$ frame, this corresponds to deflection by an angle $\sim1/\gamma_{\rm w}$, over a time scale $\Delta t_{\vert\rm w}\,\simeq\,\left(\gamma_{\rm w}^2\nu_{\rm scatt\vert\rm w}\right)^{-1}$. Accounting for time dilation $\Delta t_{\vert\rm w}\,=\,\gamma_{\rm w}(1+\mu)\Delta t$, one eventually obtains $\nu_{\rm scatt\vert\rm w}\,\sim\,\gamma_{\rm w}^{-3}l_{\rm scatt}^{-1}$. Alternatively, computing infinitesimal variations of pitch angle $\Delta\theta_{\vert\rm w}$ and $\Delta\theta$ in the respective $\mathcal R_{\rm w}$ and $\mathcal R_{\rm s}$ frames, respectively on timescales $\Delta t_{\vert\rm w}$ and $\Delta t$, one finds
\begin{equation}
  \nu_{{\rm scatt}\vert\rm w}\,\equiv\,\frac{\left\langle\Delta
    \theta_{\vert\rm w}^2\right\rangle}{2\Delta t_{\vert\rm w}}\,\simeq\,
  \frac{1}{\gamma_{\rm w}^3\left(1+\mu\right)}\frac{\left\langle\Delta
    \theta^2\right\rangle}{2\Delta t}\,.
\end{equation}
Then, the average of $\nu_{\rm scatt\vert\rm w}$ over an isotropic distribution in $\mu_{\vert\rm w}$ in the interval $[\beta,1]$ gives
\begin{equation}
  \nu_{\rm scatt\vert\rm w}\,\simeq\,\frac{3}{4\gamma_{\rm w}^3}\nu_{\rm scatt}\,,
\label{eq:nunup}
\end{equation}
which matches the previous estimate. In the following, we re-absorb for convenience the prefactor times the factor $4$ appearing in front of $\nu_{\rm scatt\vert\rm w}$ into our expression for $\nu_{\rm scatt} \equiv l_{\rm scatt}$ (determined up to a prefactor) so that $\nu_{\rm scatt}$ and $f_{\rm b} \propto \exp\left(-\nu_{\rm scatt} x\right)$. 

In terms of shock frame pitch-angle cosine, one also derives
\begin{equation}
  \theta_{\vert\rm w}\,\simeq\,\frac{1}{\gamma_{\rm w}}\sqrt{\frac{1-\mu}{1+\mu}}\,,
\end{equation}
hence
\begin{equation}
  f_{\rm b}\,\simeq\, C(p_{\vert\rm w}) \exp\left[-\nu_{\rm scatt}(p_{\vert\rm w}) x -
    \frac{1-\mu}{1+\mu}\right]\,.
  \label{eq:fb4}
\end{equation}
Using $C(p_{\vert\rm w})\,\propto\,p_{\vert\rm w}^{-s-2}$, expressing $p_{\vert\rm w}=\gamma_{\rm w}(1-\beta_{\rm w}\mu)p$ and approximating $\beta_{\rm w}\,\simeq\,-1$, one recast $f_{\rm b}$ as
\begin{align}
  f_{\rm b}(x,p)&\,\simeq\, C\left[p(1+\mu)\right]
  \exp\left\{-\nu_{\rm scatt}\left[p(1+\mu)\right]x -
    \frac{1-\mu}{1+\mu}\right\}\nonumber\\
&\,\propto\,\left(\frac{p}{p_{\rm
      m}}\right)^{-s-2}(1+\mu)^{-s-2}\nonumber\\
&\quad\quad\times\exp\left[-\nu_{\rm scatt}(p_{\rm
        m})\frac{p^{-2}(1+\mu)^{-2}}{p_{\rm m}^{-2}}x -
    \frac{1-\mu}{1+\mu}\right]\,.
  \label{eq:fb5}
\end{align}
The above introduces a pivot momentum $p_{\rm m}$ and its corresponding scattering frequency in the shock frame $\nu_{\rm m}\,\equiv\,\nu(p_{\rm m})$; $p_{\rm m}$  characterizes here the minimum momentum of the beam distribution function at a distance $\nu_{\rm m}^{-1}$ away from the shock front.

One is usually interested in power-law solutions $C(p)\propto p^{-s-2}$; however, in making comparison to PIC simulations, one must keep in mind that the extent of the power-law in the simulation is rather small, of the order of one decade, which modifies the scalings of $\beta_{\rm b}$ and $\xi_{\rm b}$ below. Hence, we will also consider exponentially suppressed powerlaw-like solutions. To be complete, note that one can also derive the above solution directly in the shock frame, by writing the pitch-angle scattering operator in terms of shock frame coordinates $p$ and $\mu$, and taking the asymptotic limit $\beta_{\rm w}\,\rightarrow\,-1$. One then finds that $f_{\rm b}$ obeys the following equation:
\begin{align}
  \mu\partial_x f_{\rm
    b}\,=\,&\frac{1}{2}\gamma_{\rm w}^3\nu_{\rm scatt\vert  w}\left[p(1+\mu)\right]
\biggl\{
  (1+\mu)^2\partial_\mu(1-\mu^2)\partial_\mu\nonumber\\
&\quad\quad + 2(1+\mu)p\partial_p
  - 2(1+\mu)(1-\mu^2)p\partial_\mu\partial_p \nonumber\\
&\quad\quad +
  (1-\mu^2)p^2\partial_p^2\biggr\}f_{\rm b}\,.
\end{align}
One can verify that the differential operator commutes with $p(1+\mu)=p_{\vert\rm w}/\gamma$, as expected since this operator is nothing but the pitch angle scattering operator in $\mu_{\vert\rm w}$ at constant $p_{\vert\rm w}$, and that $f_{\rm b}$ as given in Eq.~(\ref{eq:fb5}) is a solution to the above equation. 

\begin{figure}
\centering\includegraphics[width=0.48\textwidth]{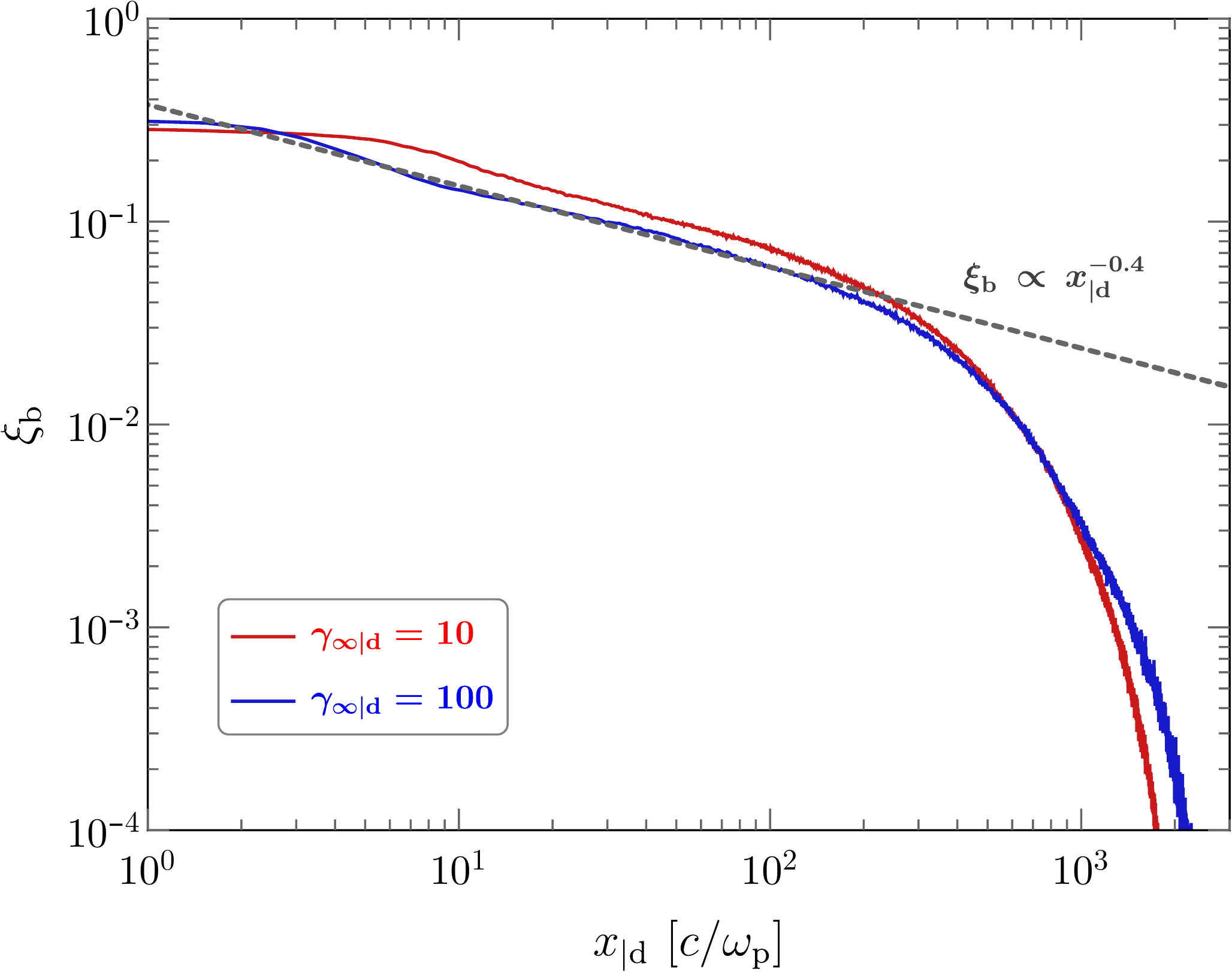}
 \caption{Behavior of $\xi_{\rm b}$ as a function of distance $x_{\vert\rm d}$, in PIC simulations with $\gamma_{\infty\vert\rm d}\,=\,10$ (red), and $\gamma_{\infty\vert\rm d}\,=\,100$ (blue). The dashed gray line shows the best-fitting power-law, $\xi_{\rm b}\,\propto\,x_{\vert\rm d}^{-0.4}$.
  \label{fig:xibvsx} }
\end{figure}

The beam distribution function can be normalized through the parameter $\xi_{\rm b}$, which we recall is defined as the ratio of the pressure of suprathermal particles to the incoming asymptotic momentum flux at infinity~\cite{L1,pap1,pap2}, 
\begin{equation}
  \xi_{\rm b}\,\equiv\,\frac{1}{\gamma_\infty^2 n_\infty m}\,\frac{2\pi}{3}\int {\rm d}p{\rm d}\mu\,p^3\,f_{\rm b}\,.
\label{eq:xib}
\end{equation}
At distances $x\,\gg\,\nu_{\rm m}^{-1}$, and in the limit in which the distribution function can be considered as an unbounded power-law in momentum of index $-s-2$, one finds a power-law behavior for $\xi_{\rm b}(x)$,
\begin{equation}
\xi_{\rm b}(x)\,\approx\,\xi_{\rm b}(x_{\rm m})\left(x \nu_{\rm m}\right)^{\frac{2-s}{2}}\,,
\label{eq:xib2}
\end{equation}
and
\begin{align}
f_{\rm b}(x,p)\,\simeq\,&\frac{3\xi_{\rm b}(x_{\rm m})\gamma_\infty^2n_\infty m}{10 \pi\Gamma[(s-2)/2]}\,\left(\frac{p(1+\mu)}{p_{\rm m}}\right)^{-s-2}\nonumber\\
&\quad\quad\times
\exp\left[-\nu_{\rm
        m}\frac{p^{-2}(1+\mu)^{-2}}{p_{\rm m}^{-2}}x -
    \frac{1-\mu}{1+\mu}\right]\,.
  \label{eq:fb6}
\end{align}
If $p_{\rm m}\,\sim\,\gamma_\infty m$, then an order of magnitude for $\nu_{\rm m}^{-1}$ is $\epsilon_B^{-1}\,\gamma_{\rm w}\,c/\omega_{\rm p}$, or about $10-10^2\,c/\omega_{\rm p}$ in the shock vicinity, where $\epsilon_B\,\sim\,0.01-0.1$ and $\gamma_{\rm w}$ is of the order of a few. 

Figure~\ref{fig:xibvsx} plots the dependence $\xi_{\rm b}(x)$ observed in our two PIC simulations for shock Lorentz factors $\gamma_{\infty\vert\rm d}\,=\,10$ and $100$. The overlaid law $\xi_{\rm b}(x)\,\propto\,x^{-0.4}$ (dashed line) confirms that $\xi_{\rm b}(x)$ indeed follows a power-law scaling over the first $\sim\,300\,c/\omega_{\rm p}$, before turning over into an exponentially suppressed behavior. This length scale depends directly on the integration time of the simulation, $t_{\rm max\vert d}=3600\,\omega_{\rm p}^{-1}$ for $\gamma_{\infty \vert \rm d}=10$ and to $t_{\rm max\vert d}=6900\,\omega_{\rm p}^{-1}$ for $\gamma_{\infty \vert \rm d}=100$. The slight difference between the observed spatial power-law and that predicted above, $\xi_{\rm b}\,\propto\,x^{-0.1}$ for $s=2.2$, may be attributed to the difference between the spectrum of accelerated particles in the simulation and a pure powerlaw, and/or possibly to the fact that $\gamma_{\rm p}$ is evolving in this region in the numerical simulation.  Repeating the above calculation of $\xi_{\rm b}$ for such a power-law with exponential suppression an order of magnitude above the injection threshold, one indeed finds a steeper power-law for $\xi_{\rm b}$, which transits into an exponentially suppressed dependence further away.  Note that the theoretical profiles that we have derived ignore the evolution of $\epsilon_B$, which, although slow, may impact further the power-law behavior.

\begin{figure}
\includegraphics[width=0.48\textwidth]{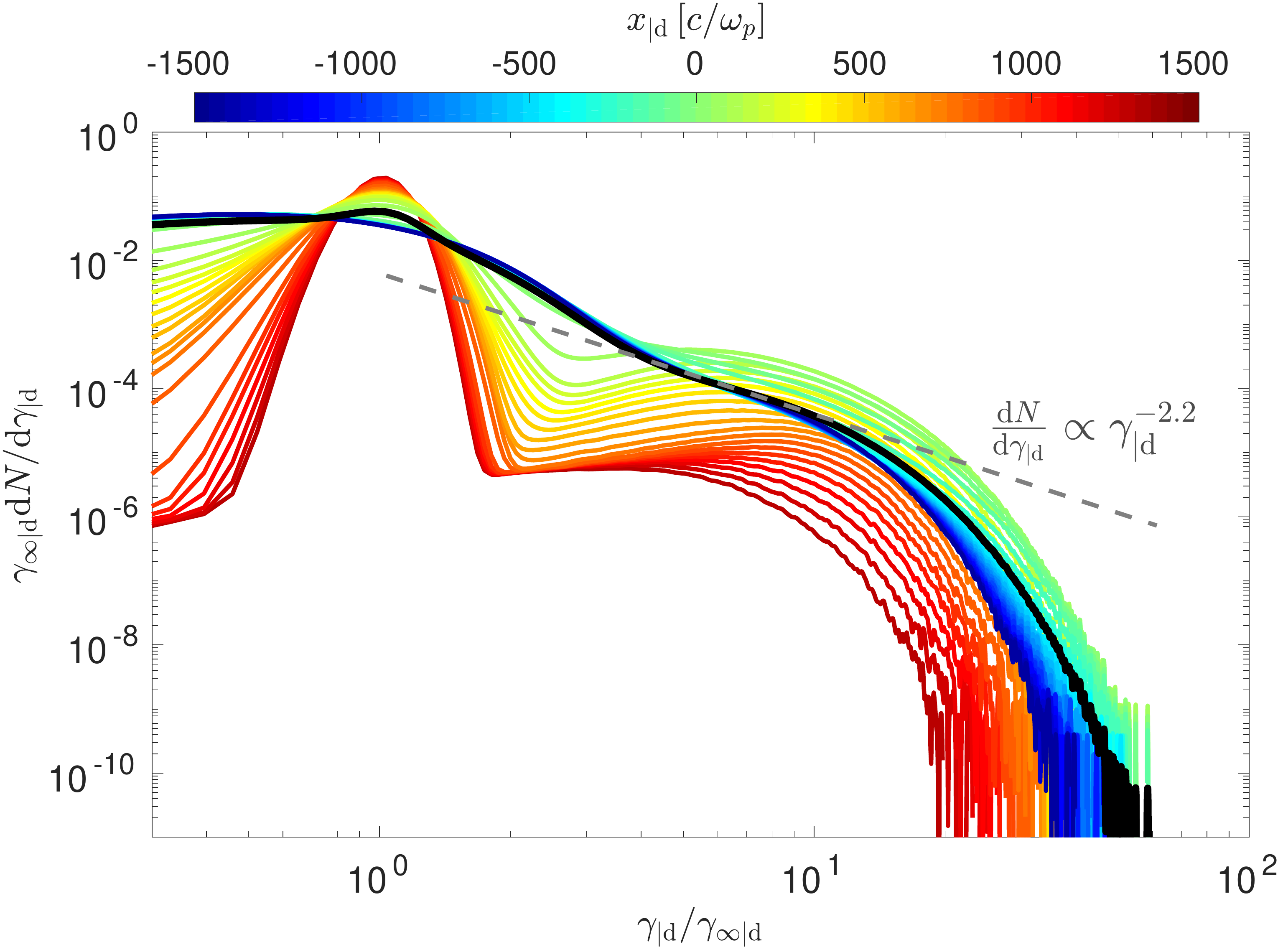}\\
\includegraphics[width=0.48\textwidth]{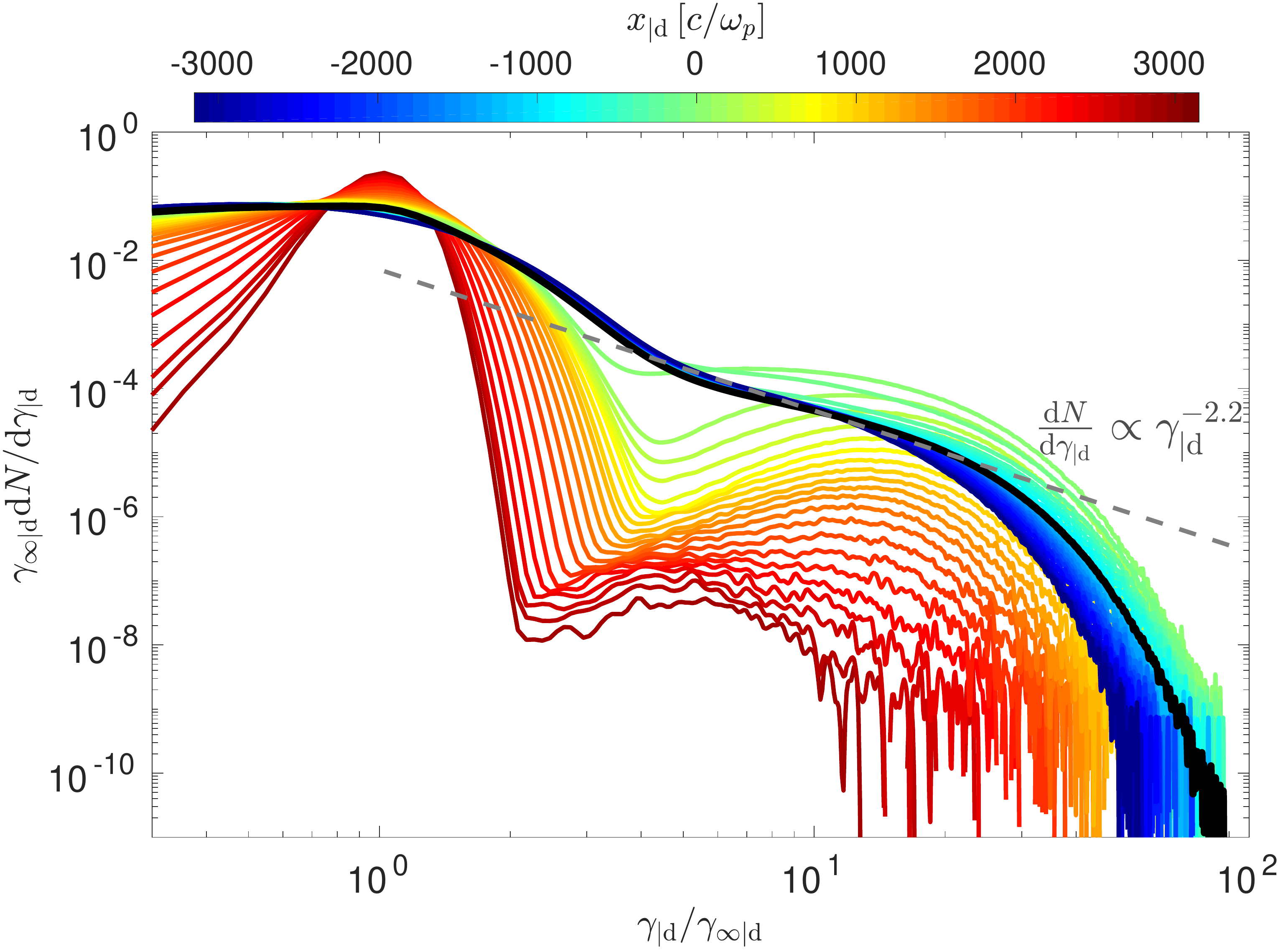}
 \caption{All-particle spectrum extracted from PIC simulations with $\gamma_{\infty\vert\rm d}\,=\,10$ (top panel) and $\gamma_{\infty\vert\rm d}\,=\,100$ (bottom panel) at various positions $x_{\vert\rm d}$ (relative to the shock front) in the simulation box, as indicated by the color bar, over a window of length $300c/\omega_{\rm p}$.  The black line plots the spectrum integrated over the full domain, while the dashed line plots the expected power-law, ${\rm d}N/{\rm d}\gamma_{\vert\rm d} \,\propto\,\gamma_{\vert\rm d}^{-2.2}$, expected in 3D momentum space~\cite{1998PhRvL..80.3911B,2000ApJ...542..235K,2001MNRAS.328..393A,2003ApJ...589L..73L, 2005PhRvL..94k1102K}.
  \label{fig:spec} }
\end{figure}

For reference, we plot in Fig.~\ref{fig:spec} the all-particle spectra ${\rm d}N/{\rm d}\gamma_{\vert\rm d}$ for our two reference PIC simulations, as integrated over the transverse dimension and over a box of length $300\,c/\omega_{\rm p}$ along $x$, centered at various positions $x_{\vert\rm d}$ as indicated. This figure shows how the spectrum evolves in the precursor. In particular, the expected (in 3D momentum space) powerlaw ${\rm d}N/{\rm d}\gamma_{\vert\rm d}\,\propto\,\gamma_{\vert\rm d}^{-2.2}$~\cite{1998PhRvL..80.3911B,2000ApJ...542..235K,2001MNRAS.328..393A,2003ApJ...589L..73L, 2005PhRvL..94k1102K,2006ApJ...645L.129L,2006ApJ...650.1020N,2015SSRv..191..519S} is recovered in the steady-state regime, in the near precursor and downstream.   

The 2D distribution function, which is needed for a proper comparison to PIC simulations, can be obtained by replacing the scattering operator according to:  $\partial_{\mu_{\vert\rm w}}(1-\mu_{\vert\rm w}^2)\partial_{\mu_{\vert\rm w}}\,\rightarrow\,\partial_{\theta_{\vert\rm w}}\partial_{\theta_{\vert\rm w}}$. One then finds a similar distribution function, up to the substitution $(1-\mu)/(1+\mu)\,\rightarrow\,(1/2)(1-\mu)/(1+\mu)$ in the exponential, and of course $-s-2\,\rightarrow\,-s-1$. The power-law behavior for $\xi_{\rm b}(x)$, however, remains unchanged.

Returning to 3D, the number density of suprathermal particles,
\begin{equation}
n_{\rm b}\,=\,2\pi\int{\rm d}p{\rm d}\mu\,p^2\, f_{\rm b}\,,
\end{equation}
follows a power-law $n_{\rm b}\,\propto\, (\nu_{\rm m}x)^{(1-s)/2}$ at large distances compared to $\nu_{\rm m}^{-1}$. Finally, the beam bulk velocity,
\begin{equation}
\beta_{\rm b}\,=\,\frac{2\pi}{n_{\rm b}}\int{\rm d}p{\rm d}\mu\,p^2\,\mu\, f_{\rm b}\,,
\label{eq:bb}
\end{equation}
evolves from  $\beta_{\rm b}\,\simeq\,-0.27$ at the shock front $x\,\rightarrow\,0$, to  $\beta_{\rm b}\,\rightarrow\,0$ at large distances $x\,\gg\,\nu_{\rm m}^{-1}$. The suprathermal particles are not fully isotropic in the downstream or shock front frame, but their average velocity in the shock frame remains well subrelativistic.

\subsection{Time-dependent regime}
It is of interest to study the distribution function of suprathermal particles in a time-dependent regime in order to carry out a meaningful comparison to PIC simulations. Such a comparison will notably provide a direct estimate of the scattering length of particles as a function of their energy, which can then be compared to our quasilinear estimate Eq.~(\ref{eq:lscfin}). To approximate this distribution function in the time-dependent regime, $t\,\ll\,\lsc$, To this goal, we assume that particles move along straight lines but can at any time interact to suffer a deflection into the opposite half-space of pitch angle -- {\it i.e.}, those with $\mu\,>\,0$ are deflected in the half-space $\mu\,<\,0$ and vice versa -- with a mean waiting time of $\lsc$. We suppose that the shock front injects ${\rm d}\dot N/{\rm d}p{\rm d}\mu$ particles (per unit transverse area of the shock front) per unit time, momentum interval and pitch angle cosine interval. We also distinguish between forward-  and backward-moving beam particles: the backward-moving particles result from the deflection of forward-moving particles. Since about half or more of the particles come back to the shock front after experiencing only one interaction, we neglect the possibility of multiple interactions in this time-dependent regime.

The forward-moving beam particles, with distribution function $f_{\rm b>}(x)$, thus correspond to the injected population that has not experienced any deflection up to distance $x$, and reads
\begin{equation}
  f_{\rm b>}(x)\,=\,\frac{1}{2\pi p^2}\frac{{\rm d}\dot N}{{\rm
      d}p{\rm d}\mu}\,e^{-x/(\mu\lsc)}\,,
\end{equation}
since $x/\mu$ indicates the time spent since injection at the shock front. Integrating over pitch-angle cosine, and assuming isotropic injection at the shock front, we derive the position dependent density
\begin{equation}
  \frac{{\rm d}n_{\rm b>}}{{\rm d}p{\rm d}x}\,=\,\frac{{\rm d}\dot N}{{\rm
      d}p}\,\Gamma\left[0,\frac{x}{\lsc}\right]\,,
\end{equation}
keeping in mind that $\lsc$ also depends on $p$. The logarithmic divergence at $x\,\rightarrow\,0$ is an artefact that results from our assumption of an infinitely thin shock front; in the following, we will regularize it as ${\rm d}n_{\rm b>}/{\rm d}p{\rm d}x\,\approx\,{\rm d}\dot N/{\rm d}p$ as $x\,\ll\,\lsc$.

The backward-moving particles, although fewer in number, play a special role, as will be shown in the following. Their distribution can also be obtained assuming straight-line trajectories. Consider such a particle at $x$ at $t$, with momentum $p$ and pitch-angle cosine $\mu$. In our approximation, this particle results from the deflection of a forward-moving particle at some point $x_1\,\geq\,x$, and some time $t_1\,\leq\,t$, which itself was emitted by the shock front at some time $t_0\,\leq\,t_1$, with pitch-angle cosine $\mu_0$ and momentum $p_0$. For simplicity, we neglect the order of unity energy gain experienced by the particle suffers during its deflection and set $p_0\,=\,p$. If $G(x,t,\mu;x_1,t_1,\mu_1)$ denotes the propagator representing rectilinear propagation without interaction, from coordinates $(x_1,t_1,\mu_1)$ to $(x,t,\mu)$, then the density of backward-moving particles can be written
\begin{align}
  \frac{{\rm d}n_{\rm b<}}{{\rm d}p{\rm d}x{\rm d}\mu}\,=\,&\frac{1}{\mu}
  \int_0^t{\rm d}t_0\int_0^1{\rm d}\mu_0\int_{t_0}^t{\rm
    d}t_1\int_0^{\infty}{\rm d}x_1 \int_0^1{\rm
    d}\mu_{1>}\nonumber\\
&\quad\quad \times G\left(x,t,\mu;x_1,t_1,\mu_{1<}\right)
  \,\frac{P(\mu_{1<};\mu_{1>})}{\lsc}\nonumber\\
&\quad\quad\times
  G\left(x_1,t_1,\mu_{1>};0,t_0,\mu_0\right)\,\frac{{\rm d}\dot
    N}{{\rm d}p{\rm d}\mu_0}\,.
\end{align}
Here, $P(\mu_{1<};\mu_{1>})$ denotes the probability of deflecting particle with incoming $\mu_{1>}$ into $\mu_{1<}$ upon interaction; we use 
$P(\mu_{1<};\mu_{1>})\,=\,\Theta\left[-\mu_{1<}\,\mu_{1>}\right]$. For ballistic transport,
\begin{align}
  G\left(x,t,\mu;x_1,t_1,\mu_{1}\right)\,=\,&e^{-\vert
      t-t_1\vert/\lsc}\,\delta\left(\mu-\mu_1\right)\nonumber\\
&\quad\quad\times
  \delta\left[t_1-t_0-\frac{x-x_1}{\mu}\right]\,.
\end{align}
Assuming isotropic injection at the shock front, we then obtain
\begin{equation}
  \frac{{\rm d}n_{\rm b<}}{{\rm d}p{\rm d}x{\rm d}\mu}\,=\,
  \left(1-\frac{x}{t}\right)\left[1-\exp\left(-\frac{
      t}{\lsc}\right)\right] \frac{{\rm d}\dot N}{{\rm d}p}
  \,\simeq\, \frac{t-x}{\lsc}\,\frac{{\rm d}\dot N}{{\rm d}p}\,,
\end{equation}
since $t\,\ll\,\lsc$ by assumption. Note that the coordinates are expressed in the shock frame; in terms of downstream frame coordinates, which are more appropriate for a direct comparison with the PIC simulation, the above becomes
\begin{equation}
  \frac{{\rm d}n_{\rm b<}}{{\rm d}p{\rm d}x{\rm d}\mu}\,\simeq\,
  \frac{1}{\gamma_{\rm d}\left(1+\beta_{\rm d}\right)}
  \frac{\left(1+\beta_{\rm d}\right)t_{\vert\rm d} - \left(
    x_{\rm\vert d}+\beta_{\rm d}t_{\rm\vert d}\right)}{\lsc}
  \,\frac{{\rm d}\dot N}{{\rm d}p}\,,
  \label{eq:dnbb}
\end{equation}
with $\beta_{\rm d}$ the velocity of the downstream relative to the shock front; in a 2D PIC simulation, $\beta_{\rm d}\,\simeq\,-1/2$. The quantity in the numerator represents the difference between the total precursor length and the distance between the shock and the particle, in the simulation rest frame. It thus indicates that the backward-moving particle density vanishes at the tip of the precursor, then increases linearly as one nears the shock front.  Quite interestingly, the above dependence of the backward-moving suprathermal particles distribution on distance offers a direct means to infer the scattering length scale of these particles from a PIC simulation. By contrast, this scattering length scale does not affect the distribution of forward-moving suprathermal particles in the time-dependent regime, since they propagate along nearly straight lines. It does control the distribution of the suprathermal particle population in the stationary state, but, as discussed above, this stationary state is limited to the lowest momenta range for a PIC simulation of reasonable duration, and is thus sensitive to the details of the turbulence in the vicinity of the shock transition.

\begin{figure}
\includegraphics[width=0.48\textwidth]{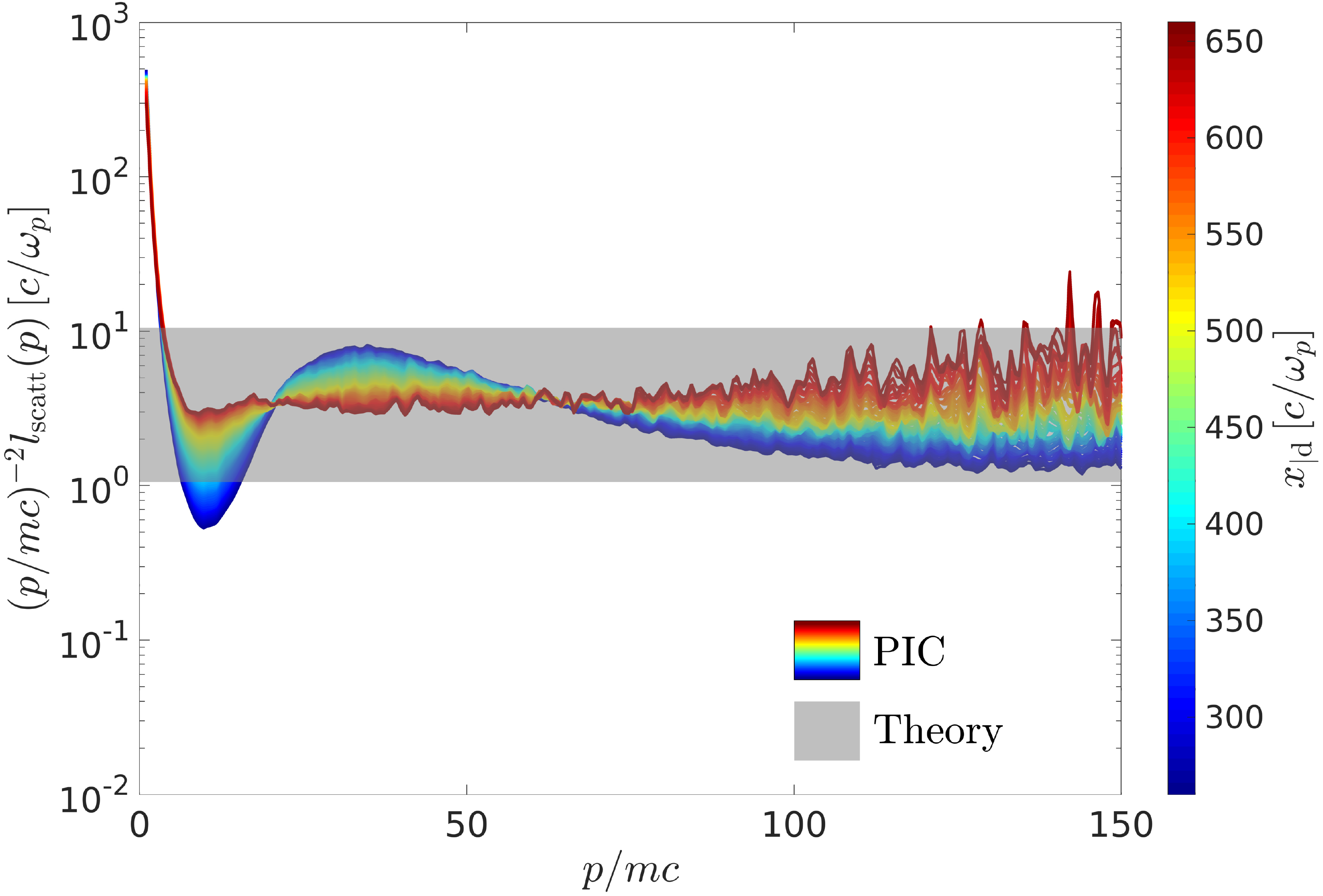}\\
\includegraphics[width=0.48\textwidth]{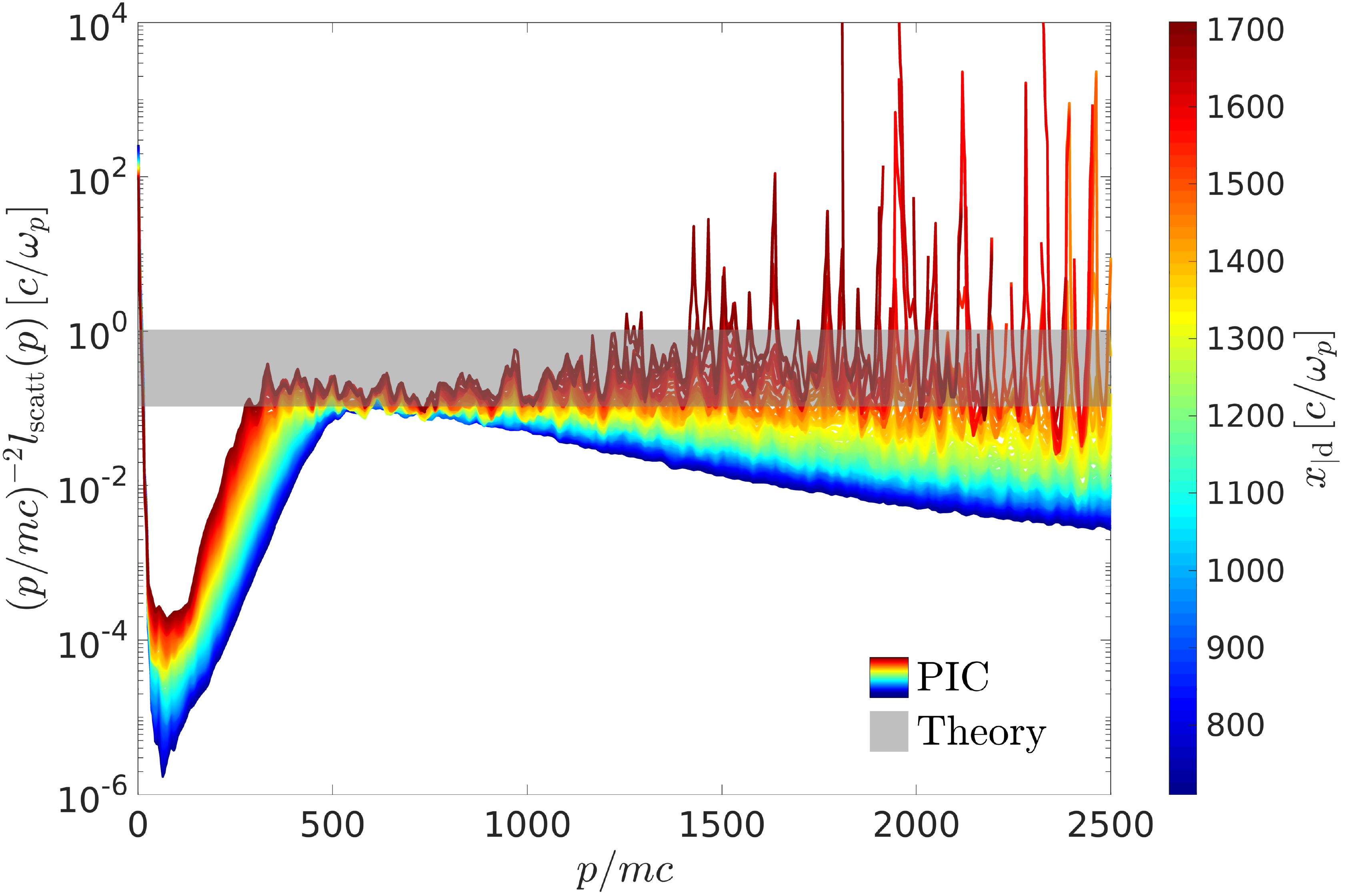}
 \caption{Comparison between our theoretical estimate of the scattering length $l_{\rm scatt}(p)\,=\,\gamma_{\rm w}\epsilon_B^{-1}(p/p_{\rm m})^2\,c/\omega_{\rm p}$ [see Eq.~(\ref{eq:lscfin})] and its measurement in PIC simulations with $\gamma_{\infty\vert\rm d}\,=\,10$ (top panel) and $\gamma_{\infty\vert\rm d}\,=\,100$ (bottom panel). The theoretical prediction includes an uncertainty error bar of a factor 3 in each direction (gray band). The numerical value is obtained through  Eq.~(\ref{eq:dnbb}) at various positions in the precursor: the values $x_{\vert\rm d}$ of interest are those corresponding to the far precursor (red colors), where the time-dependent regime applies, see text for details. The flatness of $(p/mc)^{-2}l_{\rm scatt}$ vs $p/mc$ confirms the scaling $l_{\rm scatt}\,\propto\,p^2$.
  \label{fig:lscatt10} }
\end{figure}

The two panels of Fig.~\ref{fig:lscatt10} plot the quantity $(p/p_{\rm m})^{-2}\Delta x {\rm d}N_>/{\rm d}N_<$, as measured in our PIC simulations, with $p$ the momentum of a particle, $\Delta x\,=\,x_{\rm max}-x$ the distance between the tip of the accelerated particle population and $x$, and ${\rm d}N_>,{\rm d}N_<$ the ratio between the populations of forward- to backward-moving particles at $p$ and $x$. According to Eq.~(\ref{eq:dnbb}), this provides a direct estimate of the scattering length divided by $(p/p_{\rm m})^2$. In these two panels, the color code indicates the position $x$ where the estimate is taken. At large distances to the shock (corresponding to yellow/red colors), and at large momenta, where the assumed time-dependent regime should hold, all curves nearly lie on top of each other, indicating a coherent value of the scattering length. Furthermore, these curves do not depend on $p$, indicating that the scattering length does indeed scale as $p^2$. The gray band shows the value corresponding to the theoretical estimate given in Eq.~(\ref{eq:lscfin}), with a width corresponding to an uncertainty of a factor $3$ on either side. The satisfactory agreement obtained for all (large) values of $p$, (large) values of $x$ and the different values of $\gamma$ suggests that the simple formula (\ref{eq:lscfin}) captures the leading dependencies of the scattering length in the shock precursor.

\section{Discussion -- Conclusions}\label{sec:disc}
In this third paper of a series dedicated to the physics of unmagnetized, relativistic collisionless pair shocks, we have discussed the characteristics of the suprathermal population in the shock precursor. In particular, we have provided a theoretical estimate of the scattering length of these particles on the microscopic ``Weibel'' like turbulence, taking proper account of the anisotropy of this turbulence, its growth in the precursor and its relativistic motion with respect to the shock front, see~\cite{L1,pap1,pap2}. We have obtained the formula: $\lsc(p)\,\approx\,\epsilon_B^{-1}\gamma_{\rm w}(p/p_{\rm m})^2\,c/\omega_{\rm p}$, with $p_{\rm m}\,=\,\gamma_\infty m$ the typical injection momentum of the accelerated particle distribution and $\gamma_{\rm w}$ the Lorentz factor of the turbulence relative to the shock front. We have compared this value to values extracted from dedicated large-scale PIC simulations and found satisfactory agreeement, both in terms of the prefactor and of the exponent of $p$.
Furthermore, we have characterized the distribution of the suprathermal particles in the shock precursor and deduced the spatial profile of their momentum flux. Here again, reasonable agreement is found when making comparison with PIC simlulations.

A comment is worthwhile on the $\gamma_{\rm w} \simeq \gamma_{\rm p}$ prefactor in $l_{\rm scatt}(p)$. This prefactor, which derives from the relativistic motion between the ``Weibel frame'' and the shock frame (in which the suprathermal particle beam is roughly isotropic), bears important consequences for the phenomenology of such shock waves. This is so because $l_{\rm scatt}(p)$ sets the typical value of the residence time in the $\mathcal R_{\rm s}$ frame, hence the typical value of the acceleration timescale. The residence time in the downstream plasma is expected to be $\gamma_\infty$ times shorter, because the downstream flow is not relativistic relative to the shock front. The acceleration timescale has been directly measured in long-timescale PIC simulations~\cite{2013ApJ...771...54S}. At present times, however, such simulations can only probe the early development of the power-law of accelerated particles, because $t_{\rm acc}(p)\,\propto\,p^2$ implies that the maximum energy $p_{\rm max}\,\propto\,t^{1/2}$. Furthermore, most of the accelerated particles in such simulations have gained energy in the shock vicinity. This is clearly seen in Fig.~\ref{fig:pspace}: the largest extent of the spectrum in $p_x$ of the accelerated particles is reached in the near precursor. In our model, this behavior can be ascribed to the scattering length increasing with $\gamma_{\rm w}$, and hence with the penetration depth into the precursor (since $\gamma_{\rm w}$ drops from $\gamma_\infty$ at the tip of the precursor down to $\sim1$ at $x\,\sim\,0$), which renders grazing Fermi orbits more likely in a given simulation time. This also implies that current PIC simulations fail to access the acceleration timescale in the far precursor that controls the generation of the highest-energy particles in a realistic setting.

Most likely, the transport of these highest-energy particles in the far upstream would then become dominated by an external magnetic field, even if as weak as that of the interstellar medium, because the microturbulence implies an acceleration timescale $t_{\rm acc}\,\propto\,p^2$, while a regular magnetic field guarantees $t_{\rm acc}\,\propto\,p$. Hence, in spite of the inefficient scattering implied by the relativistically moving ``Weibel'' turbulence, the acceleration timescale in the interstellar magnetic field remains short enough to ensure that synchrotron GeV photons can be produced in the early phases of a highly energetic gamma-ray burst afterglow~\cite{2013MNRAS.430.1280P}.\bigskip

\begin{acknowledgments} We acknowledge financial support from the Programme National Hautes \'Energies (PNHE) of the C.N.R.S., the ANR-14-CE33-0019 MACH project and the ILP LABEX (under reference ANR-10-LABX-63) as part of the Idex SUPER, and which is financed by French state funds managed by the ANR within the "Investissements d'Avenir" program under reference ANR-11-IDEX-0004-02. This work was granted access to the HPC resources of TGCC/CCRT under the allocation 2018-A0030407666 made by GENCI. We acknowledge PRACE for awarding us access to resource Joliot Curie-SKL based in France at TGCC Center.
\end{acknowledgments}

\bibliographystyle{apsrev4-1}

\bibliography{shock}

\begin{thebibliography}{46}%
\makeatletter
\providecommand \@ifxundefined [1]{%
 \@ifx{#1\undefined}
}%
\providecommand \@ifnum [1]{%
 \ifnum #1\expandafter \@firstoftwo
 \else \expandafter \@secondoftwo
 \fi
}%
\providecommand \@ifx [1]{%
 \ifx #1\expandafter \@firstoftwo
 \else \expandafter \@secondoftwo
 \fi
}%
\providecommand \natexlab [1]{#1}%
\providecommand \enquote  [1]{``#1''}%
\providecommand \bibnamefont  [1]{#1}%
\providecommand \bibfnamefont [1]{#1}%
\providecommand \citenamefont [1]{#1}%
\providecommand \href@noop [0]{\@secondoftwo}%
\providecommand \href [0]{\begingroup \@sanitize@url \@href}%
\providecommand \@href[1]{\@@startlink{#1}\@@href}%
\providecommand \@@href[1]{\endgroup#1\@@endlink}%
\providecommand \@sanitize@url [0]{\catcode `\\12\catcode `\$12\catcode
  `\&12\catcode `\#12\catcode `\^12\catcode `\_12\catcode `\%12\relax}%
\providecommand \@@startlink[1]{}%
\providecommand \@@endlink[0]{}%
\providecommand \url  [0]{\begingroup\@sanitize@url \@url }%
\providecommand \@url [1]{\endgroup\@href {#1}{\urlprefix }}%
\providecommand \urlprefix  [0]{URL }%
\providecommand \Eprint [0]{\href }%
\providecommand \doibase [0]{http://dx.doi.org/}%
\providecommand \selectlanguage [0]{\@gobble}%
\providecommand \bibinfo  [0]{\@secondoftwo}%
\providecommand \bibfield  [0]{\@secondoftwo}%
\providecommand \translation [1]{[#1]}%
\providecommand \BibitemOpen [0]{}%
\providecommand \bibitemStop [0]{}%
\providecommand \bibitemNoStop [0]{.\EOS\space}%
\providecommand \EOS [0]{\spacefactor3000\relax}%
\providecommand \BibitemShut  [1]{\csname bibitem#1\endcsname}%
\let\auto@bib@innerbib\@empty
\bibitem [{\citenamefont {{Bykov}}\ \emph {et~al.}(2012)\citenamefont
  {{Bykov}}, \citenamefont {{Gehrels}}, \citenamefont {{Krawczynski}},
  \citenamefont {{Lemoine}}, \citenamefont {{Pelletier}},\ and\ \citenamefont
  {{Pohl}}}]{2012SSRv..173..309B}%
  \BibitemOpen
  \bibfield  {author} {\bibinfo {author} {\bibfnamefont {A.}~\bibnamefont
  {{Bykov}}}, \bibinfo {author} {\bibfnamefont {N.}~\bibnamefont {{Gehrels}}},
  \bibinfo {author} {\bibfnamefont {H.}~\bibnamefont {{Krawczynski}}}, \bibinfo
  {author} {\bibfnamefont {M.}~\bibnamefont {{Lemoine}}}, \bibinfo {author}
  {\bibfnamefont {G.}~\bibnamefont {{Pelletier}}}, \ and\ \bibinfo {author}
  {\bibfnamefont {M.}~\bibnamefont {{Pohl}}},\ }\href {\doibase
  10.1007/s11214-012-9896-y} {\bibfield  {journal} {\bibinfo  {journal} {Sp.
  Sc. Rev.}\ }\textbf {\bibinfo {volume} {173}},\ \bibinfo {pages} {309}
  (\bibinfo {year} {2012})}\BibitemShut {NoStop}%
\bibitem [{\citenamefont {{Sironi}}\ \emph {et~al.}(2015)\citenamefont
  {{Sironi}}, \citenamefont {{Keshet}},\ and\ \citenamefont
  {{Lemoine}}}]{2015SSRv..191..519S}%
  \BibitemOpen
  \bibfield  {author} {\bibinfo {author} {\bibfnamefont {L.}~\bibnamefont
  {{Sironi}}}, \bibinfo {author} {\bibfnamefont {U.}~\bibnamefont {{Keshet}}},
  \ and\ \bibinfo {author} {\bibfnamefont {M.}~\bibnamefont {{Lemoine}}},\
  }\href {\doibase 10.1007/s11214-015-0181-8} {\bibfield  {journal} {\bibinfo
  {journal} {Sp. Sc. Rev.}\ }\textbf {\bibinfo {volume} {191}},\ \bibinfo
  {pages} {519} (\bibinfo {year} {2015})}\BibitemShut {NoStop}%
\bibitem [{\citenamefont {{Pelletier}}\ \emph {et~al.}(2017)\citenamefont
  {{Pelletier}}, \citenamefont {{Bykov}}, \citenamefont {{Ellison}},\ and\
  \citenamefont {{Lemoine}}}]{2017SSRv..207..319P}%
  \BibitemOpen
  \bibfield  {author} {\bibinfo {author} {\bibfnamefont {G.}~\bibnamefont
  {{Pelletier}}}, \bibinfo {author} {\bibfnamefont {A.}~\bibnamefont
  {{Bykov}}}, \bibinfo {author} {\bibfnamefont {D.}~\bibnamefont {{Ellison}}},
  \ and\ \bibinfo {author} {\bibfnamefont {M.}~\bibnamefont {{Lemoine}}},\
  }\href {\doibase 10.1007/s11214-017-0364-6} {\bibfield  {journal} {\bibinfo
  {journal} {Sp. Sc. Rev.}\ }\textbf {\bibinfo {volume} {207}},\ \bibinfo
  {pages} {319} (\bibinfo {year} {2017})}\BibitemShut {NoStop}%
\bibitem [{\citenamefont {{Bednarz}}\ and\ \citenamefont
  {{Ostrowski}}(1998)}]{1998PhRvL..80.3911B}%
  \BibitemOpen
  \bibfield  {author} {\bibinfo {author} {\bibfnamefont {J.}~\bibnamefont
  {{Bednarz}}}\ and\ \bibinfo {author} {\bibfnamefont {M.}~\bibnamefont
  {{Ostrowski}}},\ }\href {\doibase 10.1103/PhysRevLett.80.3911} {\bibfield
  {journal} {\bibinfo  {journal} {Physical Review Letters}\ }\textbf {\bibinfo
  {volume} {80}},\ \bibinfo {pages} {3911} (\bibinfo {year}
  {1998})}\BibitemShut {NoStop}%
\bibitem [{\citenamefont {{Kirk}}\ \emph {et~al.}(2000)\citenamefont {{Kirk}},
  \citenamefont {{Guthmann}}, \citenamefont {{Gallant}},\ and\ \citenamefont
  {{Achterberg}}}]{2000ApJ...542..235K}%
  \BibitemOpen
  \bibfield  {author} {\bibinfo {author} {\bibfnamefont {J.~G.}\ \bibnamefont
  {{Kirk}}}, \bibinfo {author} {\bibfnamefont {A.~W.}\ \bibnamefont
  {{Guthmann}}}, \bibinfo {author} {\bibfnamefont {Y.~A.}\ \bibnamefont
  {{Gallant}}}, \ and\ \bibinfo {author} {\bibfnamefont {A.}~\bibnamefont
  {{Achterberg}}},\ }\href {\doibase 10.1086/309533} {\bibfield  {journal}
  {\bibinfo  {journal} {Astrophys. J.}\ }\textbf {\bibinfo {volume} {542}},\
  \bibinfo {pages} {235} (\bibinfo {year} {2000})}\BibitemShut {NoStop}%
\bibitem [{\citenamefont {{Achterberg}}\ \emph {et~al.}(2001)\citenamefont
  {{Achterberg}}, \citenamefont {{Gallant}}, \citenamefont {{Kirk}},\ and\
  \citenamefont {{Guthmann}}}]{2001MNRAS.328..393A}%
  \BibitemOpen
  \bibfield  {author} {\bibinfo {author} {\bibfnamefont {A.}~\bibnamefont
  {{Achterberg}}}, \bibinfo {author} {\bibfnamefont {Y.~A.}\ \bibnamefont
  {{Gallant}}}, \bibinfo {author} {\bibfnamefont {J.~G.}\ \bibnamefont
  {{Kirk}}}, \ and\ \bibinfo {author} {\bibfnamefont {A.~W.}\ \bibnamefont
  {{Guthmann}}},\ }\href {\doibase 10.1046/j.1365-8711.2001.04851.x} {\bibfield
   {journal} {\bibinfo  {journal} {MNRAS}\ }\textbf {\bibinfo {volume} {328}},\
  \bibinfo {pages} {393} (\bibinfo {year} {2001})}\BibitemShut {NoStop}%
\bibitem [{\citenamefont {{Lemoine}}\ and\ \citenamefont
  {{Pelletier}}(2003)}]{2003ApJ...589L..73L}%
  \BibitemOpen
  \bibfield  {author} {\bibinfo {author} {\bibfnamefont {M.}~\bibnamefont
  {{Lemoine}}}\ and\ \bibinfo {author} {\bibfnamefont {G.}~\bibnamefont
  {{Pelletier}}},\ }\href {\doibase 10.1086/376353} {\bibfield  {journal}
  {\bibinfo  {journal} {Astrophys. J.}\ }\textbf {\bibinfo {volume} {589}},\
  \bibinfo {pages} {L73} (\bibinfo {year} {2003})}\BibitemShut {NoStop}%
\bibitem [{\citenamefont {{Keshet}}\ and\ \citenamefont
  {{Waxman}}(2005)}]{2005PhRvL..94k1102K}%
  \BibitemOpen
  \bibfield  {author} {\bibinfo {author} {\bibfnamefont {U.}~\bibnamefont
  {{Keshet}}}\ and\ \bibinfo {author} {\bibfnamefont {E.}~\bibnamefont
  {{Waxman}}},\ }\href {\doibase 10.1103/PhysRevLett.94.111102} {\bibfield
  {journal} {\bibinfo  {journal} {Physical Review Letters}\ }\textbf {\bibinfo
  {volume} {94}},\ \bibinfo {eid} {111102} (\bibinfo {year}
  {2005})}\BibitemShut {NoStop}%
\bibitem [{\citenamefont {{Lemoine}}\ \emph {et~al.}(2006)\citenamefont
  {{Lemoine}}, \citenamefont {{Pelletier}},\ and\ \citenamefont
  {{Revenu}}}]{2006ApJ...645L.129L}%
  \BibitemOpen
  \bibfield  {author} {\bibinfo {author} {\bibfnamefont {M.}~\bibnamefont
  {{Lemoine}}}, \bibinfo {author} {\bibfnamefont {G.}~\bibnamefont
  {{Pelletier}}}, \ and\ \bibinfo {author} {\bibfnamefont {B.}~\bibnamefont
  {{Revenu}}},\ }\href {\doibase 10.1086/506322} {\bibfield  {journal}
  {\bibinfo  {journal} {Astrophys. J.}\ }\textbf {\bibinfo {volume} {645}},\
  \bibinfo {pages} {L129} (\bibinfo {year} {2006})}\BibitemShut {NoStop}%
\bibitem [{\citenamefont {{Niemiec}}\ \emph {et~al.}(2006)\citenamefont
  {{Niemiec}}, \citenamefont {{Ostrowski}},\ and\ \citenamefont
  {{Pohl}}}]{2006ApJ...650.1020N}%
  \BibitemOpen
  \bibfield  {author} {\bibinfo {author} {\bibfnamefont {J.}~\bibnamefont
  {{Niemiec}}}, \bibinfo {author} {\bibfnamefont {M.}~\bibnamefont
  {{Ostrowski}}}, \ and\ \bibinfo {author} {\bibfnamefont {M.}~\bibnamefont
  {{Pohl}}},\ }\href {\doibase 10.1086/506901} {\bibfield  {journal} {\bibinfo
  {journal} {ApJ}\ }\textbf {\bibinfo {volume} {650}},\ \bibinfo {pages} {1020}
  (\bibinfo {year} {2006})}\BibitemShut {NoStop}%
\bibitem [{\citenamefont {{Medvedev}}\ and\ \citenamefont
  {{Loeb}}(1999)}]{1999ApJ...526..697M}%
  \BibitemOpen
  \bibfield  {author} {\bibinfo {author} {\bibfnamefont {M.~V.}\ \bibnamefont
  {{Medvedev}}}\ and\ \bibinfo {author} {\bibfnamefont {A.}~\bibnamefont
  {{Loeb}}},\ }\href {\doibase 10.1086/308038} {\bibfield  {journal} {\bibinfo
  {journal} {ApJ}\ }\textbf {\bibinfo {volume} {526}},\ \bibinfo {pages} {697}
  (\bibinfo {year} {1999})}\BibitemShut {NoStop}%
\bibitem [{\citenamefont {{Gruzinov}}\ and\ \citenamefont
  {{Waxman}}(1999)}]{1999ApJ...511..852G}%
  \BibitemOpen
  \bibfield  {author} {\bibinfo {author} {\bibfnamefont {A.}~\bibnamefont
  {{Gruzinov}}}\ and\ \bibinfo {author} {\bibfnamefont {E.}~\bibnamefont
  {{Waxman}}},\ }\href {\doibase 10.1086/306720} {\bibfield  {journal}
  {\bibinfo  {journal} {Astrophys. J.}\ }\textbf {\bibinfo {volume} {511}},\
  \bibinfo {pages} {852} (\bibinfo {year} {1999})}\BibitemShut {NoStop}%
\bibitem [{\citenamefont {{Lemoine}}\ \emph
  {et~al.}(2019{\natexlab{a}})\citenamefont {{Lemoine}}, \citenamefont
  {{Gremillet}}, \citenamefont {{Pelletier}},\ and\ \citenamefont
  {{Vanthieghem}}}]{L1}%
  \BibitemOpen
  \bibfield  {author} {\bibinfo {author} {\bibfnamefont {M.}~\bibnamefont
  {{Lemoine}}}, \bibinfo {author} {\bibfnamefont {L.}~\bibnamefont
  {{Gremillet}}}, \bibinfo {author} {\bibfnamefont {G.}~\bibnamefont
  {{Pelletier}}}, \ and\ \bibinfo {author} {\bibfnamefont {A.}~\bibnamefont
  {{Vanthieghem}}},\ }\href {\doibase 10.1103/PhysRevLett.123.035101}
  {\bibfield  {journal} {\bibinfo  {journal} {Phys. Rev. Lett.}\ }\textbf
  {\bibinfo {volume} {123}},\ \bibinfo {pages} {035101} (\bibinfo {year}
  {2019}{\natexlab{a}})},\ \Eprint {http://arxiv.org/abs/1907.07595}
  {arXiv:1907.07595} \BibitemShut {NoStop}%
\bibitem [{\citenamefont {{Pelletier}}\ \emph {et~al.}(2019)\citenamefont
  {{Pelletier}}, \citenamefont {{Gremillet}}, \citenamefont {{Vanthieghem}},\
  and\ \citenamefont {{Lemoine}}}]{pap1}%
  \BibitemOpen
  \bibfield  {author} {\bibinfo {author} {\bibfnamefont {G.}~\bibnamefont
  {{Pelletier}}}, \bibinfo {author} {\bibfnamefont {L.}~\bibnamefont
  {{Gremillet}}}, \bibinfo {author} {\bibfnamefont {A.}~\bibnamefont
  {{Vanthieghem}}}, \ and\ \bibinfo {author} {\bibfnamefont {M.}~\bibnamefont
  {{Lemoine}}},\ }\href {\doibase 10.1103/PhysRevE.100.013205} {\bibfield
  {journal} {\bibinfo  {journal} {Phys. Rev. E}\ }\textbf {\bibinfo {volume}
  {100}},\ \bibinfo {pages} {013205} (\bibinfo {year} {2019})},\ \Eprint
  {http://arxiv.org/abs/1907.07750} {arXiv:1907.07750} \BibitemShut {NoStop}%
\bibitem [{\citenamefont {{Lemoine}}\ \emph
  {et~al.}(2019{\natexlab{b}})\citenamefont {{Lemoine}}, \citenamefont
  {{Vanthieghem}}, \citenamefont {{Pelletier}},\ and\ \citenamefont
  {{Gremillet}}}]{pap2}%
  \BibitemOpen
  \bibfield  {author} {\bibinfo {author} {\bibfnamefont {M.}~\bibnamefont
  {{Lemoine}}}, \bibinfo {author} {\bibfnamefont {A.}~\bibnamefont
  {{Vanthieghem}}}, \bibinfo {author} {\bibfnamefont {G.}~\bibnamefont
  {{Pelletier}}}, \ and\ \bibinfo {author} {\bibfnamefont {L.}~\bibnamefont
  {{Gremillet}}},\ }\href@noop {} {\bibfield  {journal} {\bibinfo  {journal}
  {Phys. Rev. E, submitted (Pap. II)}\ } (\bibinfo {year}
  {2019}{\natexlab{b}})},\ \Eprint {http://arxiv.org/abs/1907.08219}
  {arXiv:1907.08219 [astro-ph.HE]} \BibitemShut {NoStop}%
\bibitem [{\citenamefont {{Vanthieghem}}\ \emph {et~al.}(2019)\citenamefont
  {{Vanthieghem}}, \citenamefont {{Lemoine}}, \citenamefont {{Gremillet}},\
  and\ \citenamefont {{Pelletier}}}]{pap4}%
  \BibitemOpen
  \bibfield  {author} {\bibinfo {author} {\bibfnamefont {A.}~\bibnamefont
  {{Vanthieghem}}}, \bibinfo {author} {\bibfnamefont {M.}~\bibnamefont
  {{Lemoine}}}, \bibinfo {author} {\bibfnamefont {L.}~\bibnamefont
  {{Gremillet}}}, \ and\ \bibinfo {author} {\bibfnamefont {G.}~\bibnamefont
  {{Pelletier}}},\ }\href@noop {} {\bibfield  {journal} {\bibinfo  {journal}
  {Phys. Rev. E, in prep. (Pap. IV)}\ } (\bibinfo {year} {2019})}\BibitemShut
  {NoStop}%
\bibitem [{\citenamefont {{Plotnikov}}\ \emph {et~al.}(2013)\citenamefont
  {{Plotnikov}}, \citenamefont {{Pelletier}},\ and\ \citenamefont
  {{Lemoine}}}]{2013MNRAS.430.1280P}%
  \BibitemOpen
  \bibfield  {author} {\bibinfo {author} {\bibfnamefont {I.}~\bibnamefont
  {{Plotnikov}}}, \bibinfo {author} {\bibfnamefont {G.}~\bibnamefont
  {{Pelletier}}}, \ and\ \bibinfo {author} {\bibfnamefont {M.}~\bibnamefont
  {{Lemoine}}},\ }\href {\doibase 10.1093/mnras/sts696} {\bibfield  {journal}
  {\bibinfo  {journal} {MNRAS}\ }\textbf {\bibinfo {volume} {430}},\ \bibinfo
  {pages} {1280} (\bibinfo {year} {2013})}\BibitemShut {NoStop}%
\bibitem [{\citenamefont {{Sironi}}\ \emph {et~al.}(2013)\citenamefont
  {{Sironi}}, \citenamefont {{Spitkovsky}},\ and\ \citenamefont
  {{Arons}}}]{2013ApJ...771...54S}%
  \BibitemOpen
  \bibfield  {author} {\bibinfo {author} {\bibfnamefont {L.}~\bibnamefont
  {{Sironi}}}, \bibinfo {author} {\bibfnamefont {A.}~\bibnamefont
  {{Spitkovsky}}}, \ and\ \bibinfo {author} {\bibfnamefont {J.}~\bibnamefont
  {{Arons}}},\ }\href {\doibase 10.1088/0004-637X/771/1/54} {\bibfield
  {journal} {\bibinfo  {journal} {ApJ}\ }\textbf {\bibinfo {volume} {771}},\
  \bibinfo {eid} {54} (\bibinfo {year} {2013})}\BibitemShut {NoStop}%
\bibitem [{\citenamefont {{Achterberg}}\ \emph {et~al.}(2007)\citenamefont
  {{Achterberg}}, \citenamefont {{Wiersma}},\ and\ \citenamefont
  {{Norman}}}]{2007A&A...475...19A}%
  \BibitemOpen
  \bibfield  {author} {\bibinfo {author} {\bibfnamefont {A.}~\bibnamefont
  {{Achterberg}}}, \bibinfo {author} {\bibfnamefont {J.}~\bibnamefont
  {{Wiersma}}}, \ and\ \bibinfo {author} {\bibfnamefont {C.~A.}\ \bibnamefont
  {{Norman}}},\ }\href {\doibase 10.1051/0004-6361:20065366} {\bibfield
  {journal} {\bibinfo  {journal} {Astron. Astrophys.}\ }\textbf {\bibinfo
  {volume} {475}},\ \bibinfo {pages} {19} (\bibinfo {year} {2007})}\BibitemShut
  {NoStop}%
\bibitem [{\citenamefont {{Blandford}}\ and\ \citenamefont
  {{McKee}}(1976)}]{1976PhFl...19.1130B}%
  \BibitemOpen
  \bibfield  {author} {\bibinfo {author} {\bibfnamefont {R.~D.}\ \bibnamefont
  {{Blandford}}}\ and\ \bibinfo {author} {\bibfnamefont {C.~F.}\ \bibnamefont
  {{McKee}}},\ }\href {\doibase 10.1063/1.861619} {\bibfield  {journal}
  {\bibinfo  {journal} {Physics of Fluids}\ }\textbf {\bibinfo {volume} {19}},\
  \bibinfo {pages} {1130} (\bibinfo {year} {1976})}\BibitemShut {NoStop}%
\bibitem [{\citenamefont {{Wiersma}}\ and\ \citenamefont
  {{Achterberg}}(2004)}]{2004A&A...428..365W}%
  \BibitemOpen
  \bibfield  {author} {\bibinfo {author} {\bibfnamefont {J.}~\bibnamefont
  {{Wiersma}}}\ and\ \bibinfo {author} {\bibfnamefont {A.}~\bibnamefont
  {{Achterberg}}},\ }\href {\doibase 10.1051/0004-6361:20041882} {\bibfield
  {journal} {\bibinfo  {journal} {Astron. Astrophys.}\ }\textbf {\bibinfo
  {volume} {428}},\ \bibinfo {pages} {365} (\bibinfo {year}
  {2004})}\BibitemShut {NoStop}%
\bibitem [{\citenamefont {{Lyubarsky}}\ and\ \citenamefont
  {{Eichler}}(2006)}]{2006ApJ...647.1250L}%
  \BibitemOpen
  \bibfield  {author} {\bibinfo {author} {\bibfnamefont {Y.}~\bibnamefont
  {{Lyubarsky}}}\ and\ \bibinfo {author} {\bibfnamefont {D.}~\bibnamefont
  {{Eichler}}},\ }\href {\doibase 10.1086/505523} {\bibfield  {journal}
  {\bibinfo  {journal} {ApJ}\ }\textbf {\bibinfo {volume} {647}},\ \bibinfo
  {pages} {1250} (\bibinfo {year} {2006})}\BibitemShut {NoStop}%
\bibitem [{\citenamefont {{Achterberg}}\ and\ \citenamefont
  {{Wiersma}}(2007)}]{2007A&A...475....1A}%
  \BibitemOpen
  \bibfield  {author} {\bibinfo {author} {\bibfnamefont {A.}~\bibnamefont
  {{Achterberg}}}\ and\ \bibinfo {author} {\bibfnamefont {J.}~\bibnamefont
  {{Wiersma}}},\ }\href {\doibase 10.1051/0004-6361:20065365} {\bibfield
  {journal} {\bibinfo  {journal} {Astron. Astrophys.}\ }\textbf {\bibinfo
  {volume} {475}},\ \bibinfo {pages} {1} (\bibinfo {year} {2007})}\BibitemShut
  {NoStop}%
\bibitem [{\citenamefont {{Lemoine}}\ and\ \citenamefont
  {{Pelletier}}(2010)}]{2010MNRAS.402..321L}%
  \BibitemOpen
  \bibfield  {author} {\bibinfo {author} {\bibfnamefont {M.}~\bibnamefont
  {{Lemoine}}}\ and\ \bibinfo {author} {\bibfnamefont {G.}~\bibnamefont
  {{Pelletier}}},\ }\href {\doibase 10.1111/j.1365-2966.2009.15869.x}
  {\bibfield  {journal} {\bibinfo  {journal} {MNRAS}\ }\textbf {\bibinfo
  {volume} {402}},\ \bibinfo {pages} {321} (\bibinfo {year}
  {2010})}\BibitemShut {NoStop}%
\bibitem [{\citenamefont {{Lemoine}}\ and\ \citenamefont
  {{Pelletier}}(2011)}]{2011MNRAS.417.1148L}%
  \BibitemOpen
  \bibfield  {author} {\bibinfo {author} {\bibfnamefont {M.}~\bibnamefont
  {{Lemoine}}}\ and\ \bibinfo {author} {\bibfnamefont {G.}~\bibnamefont
  {{Pelletier}}},\ }\href {\doibase 10.1111/j.1365-2966.2011.19331.x}
  {\bibfield  {journal} {\bibinfo  {journal} {MNRAS}\ }\textbf {\bibinfo
  {volume} {417}},\ \bibinfo {pages} {1148} (\bibinfo {year}
  {2011})}\BibitemShut {NoStop}%
\bibitem [{\citenamefont {{Rabinak}}\ \emph {et~al.}(2011)\citenamefont
  {{Rabinak}}, \citenamefont {{Katz}},\ and\ \citenamefont
  {{Waxman}}}]{2011ApJ...736..157R}%
  \BibitemOpen
  \bibfield  {author} {\bibinfo {author} {\bibfnamefont {I.}~\bibnamefont
  {{Rabinak}}}, \bibinfo {author} {\bibfnamefont {B.}~\bibnamefont {{Katz}}}, \
  and\ \bibinfo {author} {\bibfnamefont {E.}~\bibnamefont {{Waxman}}},\ }\href
  {\doibase 10.1088/0004-637X/736/2/157} {\bibfield  {journal} {\bibinfo
  {journal} {ApJ}\ }\textbf {\bibinfo {volume} {736}},\ \bibinfo {eid} {157}
  (\bibinfo {year} {2011})}\BibitemShut {NoStop}%
\bibitem [{\citenamefont {{Kato}}(2007)}]{2007ApJ...668..974K}%
  \BibitemOpen
  \bibfield  {author} {\bibinfo {author} {\bibfnamefont {T.~N.}\ \bibnamefont
  {{Kato}}},\ }\href {\doibase 10.1086/521297} {\bibfield  {journal} {\bibinfo
  {journal} {\apj}\ }\textbf {\bibinfo {volume} {668}},\ \bibinfo {pages} {974}
  (\bibinfo {year} {2007})}\BibitemShut {NoStop}%
\bibitem [{\citenamefont
  {{Spitkovsky}}(2008{\natexlab{a}})}]{2008ApJ...673L..39S}%
  \BibitemOpen
  \bibfield  {author} {\bibinfo {author} {\bibfnamefont {A.}~\bibnamefont
  {{Spitkovsky}}},\ }\href {\doibase 10.1086/527374} {\bibfield  {journal}
  {\bibinfo  {journal} {ApJL}\ }\textbf {\bibinfo {volume} {673}},\ \bibinfo
  {pages} {L39} (\bibinfo {year} {2008}{\natexlab{a}})}\BibitemShut {NoStop}%
\bibitem [{\citenamefont
  {{Spitkovsky}}(2008{\natexlab{b}})}]{2008ApJ...682L...5S}%
  \BibitemOpen
  \bibfield  {author} {\bibinfo {author} {\bibfnamefont {A.}~\bibnamefont
  {{Spitkovsky}}},\ }\href {\doibase 10.1086/590248} {\bibfield  {journal}
  {\bibinfo  {journal} {ApJL}\ }\textbf {\bibinfo {volume} {682}},\ \bibinfo
  {pages} {L5} (\bibinfo {year} {2008}{\natexlab{b}})}\BibitemShut {NoStop}%
\bibitem [{\citenamefont {{Martins}}\ \emph {et~al.}(2009)\citenamefont
  {{Martins}}, \citenamefont {{Fonseca}}, \citenamefont {{Silva}},\ and\
  \citenamefont {{Mori}}}]{2009ApJ...695L.189M}%
  \BibitemOpen
  \bibfield  {author} {\bibinfo {author} {\bibfnamefont {S.~F.}\ \bibnamefont
  {{Martins}}}, \bibinfo {author} {\bibfnamefont {R.~A.}\ \bibnamefont
  {{Fonseca}}}, \bibinfo {author} {\bibfnamefont {L.~O.}\ \bibnamefont
  {{Silva}}}, \ and\ \bibinfo {author} {\bibfnamefont {W.~B.}\ \bibnamefont
  {{Mori}}},\ }\href {\doibase 10.1088/0004-637X/695/2/L189} {\bibfield
  {journal} {\bibinfo  {journal} {ApJL}\ }\textbf {\bibinfo {volume} {695}},\
  \bibinfo {pages} {L189} (\bibinfo {year} {2009})}\BibitemShut {NoStop}%
\bibitem [{\citenamefont {{Keshet}}\ \emph {et~al.}(2009)\citenamefont
  {{Keshet}}, \citenamefont {{Katz}}, \citenamefont {{Spitkovsky}},\ and\
  \citenamefont {{Waxman}}}]{2009ApJ...693L.127K}%
  \BibitemOpen
  \bibfield  {author} {\bibinfo {author} {\bibfnamefont {U.}~\bibnamefont
  {{Keshet}}}, \bibinfo {author} {\bibfnamefont {B.}~\bibnamefont {{Katz}}},
  \bibinfo {author} {\bibfnamefont {A.}~\bibnamefont {{Spitkovsky}}}, \ and\
  \bibinfo {author} {\bibfnamefont {E.}~\bibnamefont {{Waxman}}},\ }\href
  {\doibase 10.1088/0004-637X/693/2/L127} {\bibfield  {journal} {\bibinfo
  {journal} {ApJL}\ }\textbf {\bibinfo {volume} {693}},\ \bibinfo {pages}
  {L127} (\bibinfo {year} {2009})}\BibitemShut {NoStop}%
\bibitem [{\citenamefont {{Nishikawa}}\ \emph {et~al.}(2009)\citenamefont
  {{Nishikawa}}, \citenamefont {{Niemiec}}, \citenamefont {{Hardee}},
  \citenamefont {{Medvedev}}, \citenamefont {{Sol}}, \citenamefont {{Mizuno}},
  \citenamefont {{Zhang}}, \citenamefont {{Pohl}}, \citenamefont {{Oka}},\ and\
  \citenamefont {{Hartmann}}}]{2009ApJ...698L..10N}%
  \BibitemOpen
  \bibfield  {author} {\bibinfo {author} {\bibfnamefont {K.-I.}\ \bibnamefont
  {{Nishikawa}}}, \bibinfo {author} {\bibfnamefont {J.}~\bibnamefont
  {{Niemiec}}}, \bibinfo {author} {\bibfnamefont {P.~E.}\ \bibnamefont
  {{Hardee}}}, \bibinfo {author} {\bibfnamefont {M.}~\bibnamefont
  {{Medvedev}}}, \bibinfo {author} {\bibfnamefont {H.}~\bibnamefont {{Sol}}},
  \bibinfo {author} {\bibfnamefont {Y.}~\bibnamefont {{Mizuno}}}, \bibinfo
  {author} {\bibfnamefont {B.}~\bibnamefont {{Zhang}}}, \bibinfo {author}
  {\bibfnamefont {M.}~\bibnamefont {{Pohl}}}, \bibinfo {author} {\bibfnamefont
  {M.}~\bibnamefont {{Oka}}}, \ and\ \bibinfo {author} {\bibfnamefont {D.~H.}\
  \bibnamefont {{Hartmann}}},\ }\href {\doibase 10.1088/0004-637X/698/1/L10}
  {\bibfield  {journal} {\bibinfo  {journal} {ApJL}\ }\textbf {\bibinfo
  {volume} {698}},\ \bibinfo {pages} {L10} (\bibinfo {year}
  {2009})}\BibitemShut {NoStop}%
\bibitem [{\citenamefont {{Weibel}}(1959)}]{Weibel_1959}%
  \BibitemOpen
  \bibfield  {author} {\bibinfo {author} {\bibfnamefont {E.~S.}\ \bibnamefont
  {{Weibel}}},\ }\href {\doibase 10.1103/PhysRevLett.2.83} {\bibfield
  {journal} {\bibinfo  {journal} {Phys. Rev. Lett.}\ }\textbf {\bibinfo
  {volume} {2}},\ \bibinfo {pages} {83} (\bibinfo {year} {1959})}\BibitemShut
  {NoStop}%
\bibitem [{\citenamefont {Davidson}\ \emph {et~al.}(1972)\citenamefont
  {Davidson}, \citenamefont {Hammer}, \citenamefont {Haber},\ and\
  \citenamefont {Wagner}}]{Davidson_1972}%
  \BibitemOpen
  \bibfield  {author} {\bibinfo {author} {\bibfnamefont {R.~C.}\ \bibnamefont
  {Davidson}}, \bibinfo {author} {\bibfnamefont {D.~A.}\ \bibnamefont
  {Hammer}}, \bibinfo {author} {\bibfnamefont {I.}~\bibnamefont {Haber}}, \
  and\ \bibinfo {author} {\bibfnamefont {C.~E.}\ \bibnamefont {Wagner}},\
  }\href {\doibase 10.1063/1.1693910} {\bibfield  {journal} {\bibinfo
  {journal} {Phys. Fluids}\ }\textbf {\bibinfo {volume} {15}},\ \bibinfo
  {pages} {317} (\bibinfo {year} {1972})}\BibitemShut {NoStop}%
\bibitem [{\citenamefont {{Bret}}\ \emph {et~al.}(2004)\citenamefont {{Bret}},
  \citenamefont {{Firpo}},\ and\ \citenamefont
  {{Deutsch}}}]{2004PhRvE..70d6401B}%
  \BibitemOpen
  \bibfield  {author} {\bibinfo {author} {\bibfnamefont {A.}~\bibnamefont
  {{Bret}}}, \bibinfo {author} {\bibfnamefont {M.-C.}\ \bibnamefont {{Firpo}}},
  \ and\ \bibinfo {author} {\bibfnamefont {C.}~\bibnamefont {{Deutsch}}},\
  }\href {\doibase 10.1103/PhysRevE.70.046401} {\bibfield  {journal} {\bibinfo
  {journal} {Phys. Rev. E}\ }\textbf {\bibinfo {volume} {70}},\ \bibinfo {eid}
  {046401} (\bibinfo {year} {2004})}\BibitemShut {NoStop}%
\bibitem [{\citenamefont {{Bret}}\ \emph {et~al.}(2008)\citenamefont {{Bret}},
  \citenamefont {{Gremillet}}, \citenamefont {{B{\'e}nisti}},\ and\
  \citenamefont {{Lefebvre}}}]{2008PhRvL.100t5008B}%
  \BibitemOpen
  \bibfield  {author} {\bibinfo {author} {\bibfnamefont {A.}~\bibnamefont
  {{Bret}}}, \bibinfo {author} {\bibfnamefont {L.}~\bibnamefont {{Gremillet}}},
  \bibinfo {author} {\bibfnamefont {D.}~\bibnamefont {{B{\'e}nisti}}}, \ and\
  \bibinfo {author} {\bibfnamefont {E.}~\bibnamefont {{Lefebvre}}},\ }\href
  {\doibase 10.1103/PhysRevLett.100.205008} {\bibfield  {journal} {\bibinfo
  {journal} {PRL}\ }\textbf {\bibinfo {volume} {100}},\ \bibinfo {eid} {205008}
  (\bibinfo {year} {2008})}\BibitemShut {NoStop}%
\bibitem [{\citenamefont {{Bret}}\ \emph
  {et~al.}(2010{\natexlab{a}})\citenamefont {{Bret}}, \citenamefont
  {{Gremillet}},\ and\ \citenamefont {{Dieckmann}}}]{2010PhPl...17l0501B}%
  \BibitemOpen
  \bibfield  {author} {\bibinfo {author} {\bibfnamefont {A.}~\bibnamefont
  {{Bret}}}, \bibinfo {author} {\bibfnamefont {L.}~\bibnamefont {{Gremillet}}},
  \ and\ \bibinfo {author} {\bibfnamefont {M.~E.}\ \bibnamefont
  {{Dieckmann}}},\ }\href {\doibase 10.1063/1.3514586} {\bibfield  {journal}
  {\bibinfo  {journal} {Physics of Plasmas}\ }\textbf {\bibinfo {volume}
  {17}},\ \bibinfo {pages} {120501} (\bibinfo {year}
  {2010}{\natexlab{a}})}\BibitemShut {NoStop}%
\bibitem [{\citenamefont {{Bret}}\ \emph
  {et~al.}(2010{\natexlab{b}})\citenamefont {{Bret}}, \citenamefont
  {{Gremillet}},\ and\ \citenamefont {{B{\'e}nisti}}}]{2010PhRvE..81c6402B}%
  \BibitemOpen
  \bibfield  {author} {\bibinfo {author} {\bibfnamefont {A.}~\bibnamefont
  {{Bret}}}, \bibinfo {author} {\bibfnamefont {L.}~\bibnamefont {{Gremillet}}},
  \ and\ \bibinfo {author} {\bibfnamefont {D.}~\bibnamefont {{B{\'e}nisti}}},\
  }\href {\doibase 10.1103/PhysRevE.81.036402} {\bibfield  {journal} {\bibinfo
  {journal} {\pre}\ }\textbf {\bibinfo {volume} {81}},\ \bibinfo {eid} {036402}
  (\bibinfo {year} {2010}{\natexlab{b}})}\BibitemShut {NoStop}%
\bibitem [{\citenamefont {{Godfrey}}\ and\ \citenamefont
  {{Vay}}(2014)}]{Godfrey_2014}%
  \BibitemOpen
  \bibfield  {author} {\bibinfo {author} {\bibfnamefont {B.~B.}\ \bibnamefont
  {{Godfrey}}}\ and\ \bibinfo {author} {\bibfnamefont {J.-L.}\ \bibnamefont
  {{Vay}}},\ }\href {\doibase 10.1016/j.jcp.2014.02.022} {\bibfield  {journal}
  {\bibinfo  {journal} {J. Comput. Phys.}\ }\textbf {\bibinfo {volume} {267}},\
  \bibinfo {pages} {1} (\bibinfo {year} {2014})}\BibitemShut {NoStop}%
\bibitem [{\citenamefont {{Cole}}(1997)}]{Cole_1997a}%
  \BibitemOpen
  \bibfield  {author} {\bibinfo {author} {\bibfnamefont {J.~B.}\ \bibnamefont
  {{Cole}}},\ }\href {\doibase 10.1109/22.588615} {\bibfield  {journal}
  {\bibinfo  {journal} {IEEE Trans. Microw. Theory Tech.}\ }\textbf {\bibinfo
  {volume} {45}},\ \bibinfo {pages} {991} (\bibinfo {year} {1997})}\BibitemShut
  {NoStop}%
\bibitem [{\citenamefont {{Cole}}(2002)}]{Cole_2002}%
  \BibitemOpen
  \bibfield  {author} {\bibinfo {author} {\bibfnamefont {J.~B.}\ \bibnamefont
  {{Cole}}},\ }\href {\doibase 10.1109/TAP.2002.801268} {\bibfield  {journal}
  {\bibinfo  {journal} {IEEE Trans. Antennas Propag.}\ }\textbf {\bibinfo
  {volume} {50}},\ \bibinfo {pages} {1185} (\bibinfo {year}
  {2002})}\BibitemShut {NoStop}%
\bibitem [{\citenamefont {{K\"arkk\"ainen}}\ \emph {et~al.}(2006)\citenamefont
  {{K\"arkk\"ainen}}, \citenamefont {{Gjonaj}}, \citenamefont {{Lau}},\ and\
  \citenamefont {{Weiland}}}]{Karkkainen_2006}%
  \BibitemOpen
  \bibfield  {author} {\bibinfo {author} {\bibfnamefont {M.}~\bibnamefont
  {{K\"arkk\"ainen}}}, \bibinfo {author} {\bibfnamefont {E.}~\bibnamefont
  {{Gjonaj}}}, \bibinfo {author} {\bibfnamefont {T.}~\bibnamefont {{Lau}}}, \
  and\ \bibinfo {author} {\bibfnamefont {T.}~\bibnamefont {{Weiland}}},\ }in\
  \href@noop {} {\emph {\bibinfo {booktitle} {{Proc. International
  Computational Accelerator Physics Conference, Charmonix, France}}}}\
  (\bibinfo {year} {2006})\ pp.\ \bibinfo {pages} {35--40}\BibitemShut
  {NoStop}%
\bibitem [{\citenamefont {{Chang}}\ \emph {et~al.}(2008)\citenamefont
  {{Chang}}, \citenamefont {{Spitkovsky}},\ and\ \citenamefont
  {{Arons}}}]{2008ApJ...674..378C}%
  \BibitemOpen
  \bibfield  {author} {\bibinfo {author} {\bibfnamefont {P.}~\bibnamefont
  {{Chang}}}, \bibinfo {author} {\bibfnamefont {A.}~\bibnamefont
  {{Spitkovsky}}}, \ and\ \bibinfo {author} {\bibfnamefont {J.}~\bibnamefont
  {{Arons}}},\ }\href {\doibase 10.1086/524764} {\bibfield  {journal} {\bibinfo
   {journal} {ApJ}\ }\textbf {\bibinfo {volume} {674}},\ \bibinfo {pages} {378}
  (\bibinfo {year} {2008})}\BibitemShut {NoStop}%
\bibitem [{\citenamefont {{Sironi}}\ and\ \citenamefont
  {{Spitkovsky}}(2011)}]{2011ApJ...726...75S}%
  \BibitemOpen
  \bibfield  {author} {\bibinfo {author} {\bibfnamefont {L.}~\bibnamefont
  {{Sironi}}}\ and\ \bibinfo {author} {\bibfnamefont {A.}~\bibnamefont
  {{Spitkovsky}}},\ }\href {\doibase 10.1088/0004-637X/726/2/75} {\bibfield
  {journal} {\bibinfo  {journal} {ApJ}\ }\textbf {\bibinfo {volume} {726}},\
  \bibinfo {eid} {75} (\bibinfo {year} {2011})}\BibitemShut {NoStop}%
\bibitem [{\citenamefont {{Pelletier}}\ \emph {et~al.}(2009)\citenamefont
  {{Pelletier}}, \citenamefont {{Lemoine}},\ and\ \citenamefont
  {{Marcowith}}}]{2009MNRAS.393..587P}%
  \BibitemOpen
  \bibfield  {author} {\bibinfo {author} {\bibfnamefont {G.}~\bibnamefont
  {{Pelletier}}}, \bibinfo {author} {\bibfnamefont {M.}~\bibnamefont
  {{Lemoine}}}, \ and\ \bibinfo {author} {\bibfnamefont {A.}~\bibnamefont
  {{Marcowith}}},\ }\href {\doibase 10.1111/j.1365-2966.2008.14219.x}
  {\bibfield  {journal} {\bibinfo  {journal} {MNRAS}\ }\textbf {\bibinfo
  {volume} {393}},\ \bibinfo {pages} {587} (\bibinfo {year}
  {2009})}\BibitemShut {NoStop}%
\bibitem [{\citenamefont {{Plotnikov}}\ \emph {et~al.}(2011)\citenamefont
  {{Plotnikov}}, \citenamefont {{Pelletier}},\ and\ \citenamefont
  {{Lemoine}}}]{2011A&A...532A..68P}%
  \BibitemOpen
  \bibfield  {author} {\bibinfo {author} {\bibfnamefont {I.}~\bibnamefont
  {{Plotnikov}}}, \bibinfo {author} {\bibfnamefont {G.}~\bibnamefont
  {{Pelletier}}}, \ and\ \bibinfo {author} {\bibfnamefont {M.}~\bibnamefont
  {{Lemoine}}},\ }\href {\doibase 10.1051/0004-6361/201117182} {\bibfield
  {journal} {\bibinfo  {journal} {Astron. Astrophys.}\ }\textbf {\bibinfo
  {volume} {532}},\ \bibinfo {eid} {A68} (\bibinfo {year} {2011})}\BibitemShut
  {NoStop}%
\end{thebibliography}%

\end{document}